# On the Minimal Theory of Consciousness Implicit in Active Inference

Christopher J. Whyte[1,2,3], Andrew W. Corcoran[1], Jonathan Robinson[1], Ryan Smith[4], Rosalyn J. Moran[5], Thomas Parr[6], Karl J. Friston[7], Anil K. Seth[8,9], Jakob Hohwy[1*]

**Author Affiliations**

[1]Monash Centre for Consciousness & Contemplative Studies, Monash University, Melbourne, Australia
[2]Brain and Mind Centre, University of Sydney, Sydney, Australia
[3]Centre for Complex Systems, University of Sydney, Sydney, Australia
[4]Laureate Institute for Brain Research, Tulsa, OK, USA
[5]Department of Neuroimaging, King's College London, London, United Kingdom
[6]Nuffield Department of Clinical Neurosciences, University of Oxford, Oxford, UK
[7]Queen Square Institute of Neurology, University College London, UK
[8]Sussex Centre for Consciousness Science and Department of Informatics, University of Sussex, Brighton, UK
[9]Program on Brain, Mind, and Consciousness, Canadian Institute for Advanced Research, Toronto, Canada
[*]Corresponding author: jakob.hohwy@monash.edu

**Abstract**

The multifaceted nature of subjective experience poses a challenge to the study of consciousness. Traditional neuroscientific approaches often concentrate on isolated facets, such as perceptual awareness or the global state of consciousness and construct a theory around the relevant empirical paradigms and findings. Theories of consciousness are, therefore, often difficult to compare; indeed, there might be little overlap in the phenomena such theories aim to explain. Here, we take a different approach: starting with active inference, a first principles framework for modelling behaviour as (approximate) Bayesian inference, and building up to a minimal theory of consciousness, which emerges from the shared features of computational models derived under active inference. We review a body of work applying active inference models to the study of consciousness and argue that there is implicit in all these models a small set of theoretical commitments that point to a minimal (and testable) theory of consciousness.

## 1. Introduction

Conscious experience is heterogeneous and multifaceted. At first pass, the scientific study of consciousness can be divided into three related (but in practice, largely independent) research programs: the study of contents, state, and self (Seth, 2021). The contents of consciousness are the qualities or elements within experience that an agent is conscious of (e.g., the image of a red rose against a green background, or the aroma of freshly brewed coffee). Contents are studied by controlling for physical stimulus attributes and the overall state of consciousness (such as drowsiness), while varying subjective percepts (Baars, 2002). Conscious organisms also have different global states of consciousness, which are often assessed behaviourally (e.g., through the Glasgow Coma Scale; Teasdale et al., 2014) and are crucial to assess patients with disorders of consciousness. Such states include the vegetative state, various states of sleep, normal waking states, and perhaps states like delirium or psychedelia (for discussion, see Bayne et al 2016). Consciousness (at least in humans) is typically also accompanied by some form of minimal and/or narrative self-awareness (Gallagher, 2000) along with experiences of embodiment, selfhood and personhood (Ciaunica et al., 2022; Seth, 2013; Seth & Tsakiris, 2018).

Most neuroscientific theories of consciousness take some subset of these phenomena as their explanatory target and construct a theory around the relevant empirical paradigms and findings (Seth & Bayne, 2022). For example, global workspace theory (Baars, 2005; Baars et al., 2013), and its contemporary incarnation, global neuronal workspace theory (Dehaene et al., 2011; Mashour et al., 2020), were constructed around the method of contrastive analysis, which treats the awareness of perceptual contents as the dependent variable; that is, varying whether the participant is conscious of some particular content evoked by a stimulus. The global state of consciousness was initially regarded as a background condition (Dehaene et al., 2006), making global neuronal workspace theory chiefly a theory of conscious contents. Since then, the theory has also been applied to experiments that manipulate the global state of consciousness through anaesthesia (for review see Mashour et al., 2020). However, the initial assumption that conscious state is a background condition for awareness of contents, rather than a contextual construct that constrains content, is arguably still present in contemporary versions of the theory (see Bayne et al., 2016; Bayne & Carter, 2018). Higher order theories (Brown et al., 2019; Fleming, 2020; Lau & Rosenthal, 2011) are, likewise, chiefly theories of how contents become conscious and are so far silent on the relationship between the contents and state of consciousness. Similarly, integrated information theory (Albantakis et al., 2023; Oizumi et al., 2014; Tononi et al., 2016), another leading theory of consciousness, was developed with the explicit aim of solving the hard problem of consciousness (i.e., explaining why some physical structures generate subjective experience and others do not). Integrated information theory treats consciousness in wholly intrinsic terms, thereby downplaying the role of overt behaviour, which arguably plays a major role in shaping not only what we are conscious of but also in determining the qualitative character of conscious contents (O'Regan & Noë, 2001; Seth, 2014). Other theories, such as the Self-model Theory of Subjectivity (Metzinger, 2004) or the Projective Consciousness model (Rudrauf et al., 2017), focus on explaining the (seeming) presence of a self or first person perspective. There are still other theories of consciousness beyond those cited above, many of which privilege specific explanatory targets and methodological approaches. This state of affairs thus poses a double challenge: not only are theories of consciousness difficult to arbitrate between — on the basis of empirical evidence (Yaron et al., 2022) — it is also sometimes not clear whether such theories are aiming to explain the same empirical data (Seth & Bayne, 2022).

Instead of starting with one or more specific properties of consciousness, as our primary explanatory target, in this paper we start with active inference, a framework for modelling adaptive behaviour as (approximate) Bayesian inference and ask whether we can build toward a theory of consciousness. Active inference was developed in the context of theories of predictive coding (Rao & Ballard, 1999; Srinivasan et al., 1982) and the 'Helmholtz machine' (Dayan et al., 1995) that framed brain function as inference, drawing from advances in variational inference and message passing in statistics (Beal, 2003; Wainwright & Jordan, 2008; Winn & Bishop, 2005). The novelty of the framework lies not in providing unique predictions. Indeed, many of the alternative but domain-specific normative modelling frameworks in computational neuroscience and related disciplines are consistent with, or equivalent to, active inference, but tend to apply to a narrower range of contexts (Da Costa, Sajid, et al., 2020; Sajid, Ball, et al., 2021; Sajid, Da Costa, et al., 2021). Instead, the novelty of active inference lies in its generality. An extraordinarily diverse suite of behaviours can be modelled under the framework through the minimisation of the same free energy functionals (Parr et al., 2022): namely, variational free energy and expected free energy. Indeed, the active inference framework has been applied to phenomena across the cognitive and neural sciences in areas as diverse as visual search (Cullen et al., 2020; Mirza et al., 2018; Parr et al., 2021) and the comprehension and generation of language (Friston, Parr, et al., 2020; Friston, Sajid, et al., 2020). Because these phenomena all emerge when minimising the same objective functions — which can be decomposed into a small set of interpretable (quasi teleological) terms — active inference allows us to expose commonalities and differences across diverse phenomena of interest that may be obscured in a less general modelling framework. Thus, we will argue that active inference, precisely because it is not a theory of consciousness *per se*, is able to do justice to the diverse properties of conscious experience (c.f. Hohwy & Seth, 2020; Vilas et al., 2021). The argument is then that, by beginning with a general framework for modelling adaptive behaviour, the contours of a theory of consciousness will nevertheless emerge naturally, as individual conscious phenomena (i.e., canonical paradigms in consciousness science) begin to be explained by a singular formal (i.e. mathematical) formulation.

Practically, a natural first step (which is already very much underway) — in the construction of a minimal theory of consciousness from active inference — is to deploy the framework to model individual phenomena that are paradigmatic in the field of consciousness science. Having developed such models, one can then explore their computational properties and behaviour, and map variables and parameters in each model to phenomenological aspects of the specific conscious processes being modelled. Each model then becomes a building block for the theory, which grows as more models accumulate. Through this incremental process, a fully-fledged active inference theory of consciousness may gradually take shape. A key component of this process therefore lies in identifying the computational properties common among models that have both explanatory power and systematicity across the diverse phenomena within consciousness science (Atkinson et al., 2000; Seth, 2009). Importantly, in line with the initially minimal nature of the theory, we will restrict ourselves to discussing well-studied paradigms within the neuroscience of consciousness and refrain from speculating about more general functions of consciousness (including potential evolutionary origins) beyond these experimental tasks. In effect, our approach is not so much about the emergence of consciousness (see, for example, (Fleming & Michel, 2024)) but about the properties of a system that exhibits phenomena associated with consciousness."

This paper aims to speak to two key audiences which are, at present, largely independent of one another; experimental neuroscientists and cognitive scientists working on consciousness, and neuronal modellers and theorists working within the active inference framework. For the experimental community we wish to provide a precise but accessible guide to active inference and its relationship to consciousness; with particular emphasis on the theory's empirical commitments. For the community of theorists working on active inference and not currently working on consciousness we wish to highlight both the ready applicability of the framework to consciousness and the areas in need of further theoretical and formal development.

We begin in **sections 2-3** by giving a brief introduction to the active inference modelling framework highlighting the decompositions of variational free energy and expected free energy into a small set of teleologically meaningful terms. We also introduce general features of the generative model architectures that underwrite active inference and their potential neural implementation. These two sections are the most technical and may be skipped by readers already familiar with the formalism of active inference. We encourage readers unfamiliar with active inference to persevere through these opening sections. Active inference is, at bottom, a mathematical framework and to discuss it with precision we must use the relevant formalism. **Section 4** then surveys the modelling literature relevant to consciousness science and presents a novel extension of an existing model that allows active inference to contact research on the global state of consciousness. Finally, in **section 5** we argue that implicit in all these models is a set of assumptions and theoretical commitments that, once made explicit, entail a minimal but empirically productive theory of consciousness. To make the relationship between the minimal theory and experiment as clear as possible we follow recent work in the philosophy of consciousness science (Negro, 2024) and discuss the theory and its empirical commitments through the lens of Lakatos' account of 'scientific research programs' (Lakatos, 1968).

Finally, we note that the primary aim of this paper is didactic rather than polemic. That is, we will not argue for the inadequacy of other competing theories of consciousness and then offer active inference as a replacement. Rather we aim to provide a positive account of the (minimal) theory of consciousness implicit in the active inference framework. We are of the opinion that the most productive way forward is to spell out, in as precise a manner as possible, the structure of the theory, and its relationship to empirical data in the service of facilitating empirically driven theory comparison.

## 2. Variational and expected free energy

An organism that maintains its bodily integrity must be able to stay within the narrow range of states that are consistent with its existence (e.g., for mammals, maintaining a relatively constant internal body temperature); this entails that an organism will spend the majority of its time in a relatively restricted set of characteristic states (Friston, 2013; Tschantz et al., 2020). Under active inference, this is modelled by interpreting the organism's phenotype as a generative model of its niche (i.e. a model of how its sensory input is generated), which assigns a high probability to the observations associated with frequently occupied states; i.e., the states that are characteristic of the kind of organism in question (Corcoran & Hohwy, 2018; Ramstead et al., 2020; Ramstead et al., 2018, 2020, 2021). The perception-action cycle is, therefore, cast as an optimisation problem, where the primary objective function being extremised — variational free energy — stands in for the (log) evidence for a generative model of how an organism's observations are generated (Da Costa, Parr, et al., 2020). Crucially, the generative model entails a prior belief that action sequences are more plausible when they minimise expected free energy. This means there are two objective functions to consider. Specifically, (negative) variational free energy is a computationally tractable lower bound on log model evidence, and (negative) expected free energy can be thought of as approximating expected log model evidence (Parr & Friston, 2018a). Minimising variational free energy equips the organism with an approximation to model evidence that can be leveraged to infer hidden states of the world and learn various statistics of the environment. Beliefs in this model can then be used to select actions that minimise expected free energy and thus keep the agent in phenotypically characteristic states consistent with its continued survival. This licences an interpretation of behaviour as self-evidencing (Hohwy, 2016, 2020, 2021), where organisms take actions that maximise the evidence for their model of the world. The quantities that go into this self-evidencing process are crucial and we will examine each in turn.

Variational free energy (**eq.1**) is the expected difference, under the approximate posterior ($q$), between the log of the approximate posterior and the log of the generative model ($p$) — i.e., the joint probability of observations ($o$) and their causes — or hidden states ($s$). The approximate posterior reflects the inferred hidden state of the world given sensory data.

$$
\begin{aligned}
F &= \mathrm{E}_{q(s)}[\ln q(s) - \ln p(o,s)] \qquad (1) \\
&= \underbrace{D_{KL}[q(s)||p(s|o)]}_{relative\ entropy} + \underbrace{(-\ln p(o))}_{surprise} \\
&= \underbrace{D_{KL}[q(s)||p(s)]}_{complexity} - \underbrace{\mathrm{E}_{q(s)}[\ln p(o|s)]}_{accuracy}
\end{aligned}
$$

Variational free energy admits two key decompositions, each of which highlights complementary perspectives on its essential properties. Line two of equation 1 uses the product rule of probability and the fact that model evidence does not depend on hidden states (allowing us to drop the expectation operator in the second term) to decompose variational free energy into two terms: the relative entropy or KL divergence between the approximate posterior and the true posterior, and the surprise or negative model evidence (i.e. the probability of sensory data averaged over hidden states). The KL divergence is always greater than or equal to zero, so variational free energy is always greater than or equal to surprise (a.k.a., surprisal or self information). Thus, variational free energy is an upper bound on surprise, and will be identical to surprise when the approximate posterior matches the true posterior.

This first decomposition is didactically useful in understanding the properties of variational free energy, but it is not useful for describing how variational free energy is minimised. This is because the value of the true posterior is computationally intractable and therefore cannot be known by the agent. In the third line we decompose variational free energy into two terms, complexity and accuracy, each of which are computationally tractable. Complexity is the KL divergence between the approximate posterior over states and the prior over states (i.e. beliefs about hidden states prior to receiving sensory data) and scores the difference between the prior and the approximate posterior. One can think of this as a regularisation term on the magnitude of belief updating. A large change between the prior and approximate posterior results in a high complexity. Accuracy is the expected log likelihood of observations (i.e. the probability of current observations under each hidden state). Minimising variational free energy therefore requires organisms to trade off between maximising accuracy and minimising complexity. This is a crucial point. Sensory input is noisy and ambiguous; given noisy data it is always possible to increase the accuracy by shifting the approximate posterior to fit current data. However, constantly making large changes to the posterior greatly reduces the ability of the model to generalise to new observations. Without penalising for large (Bayesian) belief updates, a maximally "accurate" model will over-fit to noise and be in need of constant revision (Sengupta et al., 2013). Minimising variational free energy, therefore, ensures that agents are equipped with a generalisable explanation for — or model of — their sensed world.

In virtue of being self-organising creatures, agents must select actions that, on average, minimise variational free energy (Friston, Rigoli, Ognibene, Mathys, Fitzgerald & Pezzulo, 2015; Parr & Friston, 2019). Purely reactive actions such as reflex arcs can be formulated in terms of variational free energy minimisation by taking actions that bring about observations consistent with a (homeostatic or proprioceptive) set point encoded in prior beliefs (Buckley et al., 2017; Tschantz et al., 2022). However, more sophisticated actions or action sequences (i.e., policies) require some form of counterfactual computation over future observations (i.e., planning). This planning rests on expected free energy (**eq. 2**), which treats the observable consequences of a policy as random variables (because they have yet to be observed). Expected free energy takes a weighted sum over expected observations to approximate the expected outcome of actions under each policy; hence 'expected' free energy. One can think of variational free energy as a special case of expected free energy that pertains to the present; when observations are known and do not depend on future actions.

Thus, according to active inference, action selection, like perception, is a process of (planning as) inference. Crucially, instead of inferring the hidden states that maximise the probability of sensory outcomes, agents must infer their most probable course of action (Friston, et al 2017), where actions can include everything from overt bodily movements, such as saccades, to covert mental actions, such as the direction of attention. This scheme reverses the usual framework for modelling action selection, instead of searching for the action that will reach some preferred state, it starts by assuming that the agent will achieve their preferred state and then infers the most probable course of action to get there (c.f. Millidge et al., 2020). The most probable course of action is the policy that minimises the expected free energy of plausible policies.

Line two of **eq. 2** shows the decomposition of expected free energy into its most intuitive component parts: risk, ambiguity, and novelty. As the derivation is somewhat involved we refer interested readers to the appendix of (Da Costa, Parr, et al., 2020).

$$\begin{aligned} G_\pi = & \ \mathrm{E}_{\overline{\mathrm{q}}}[\ln q(s, \boldsymbol{A}|\pi) - \ln p(o, s, \boldsymbol{A}|\pi)] \qquad (2) \\ \approx & \underbrace{D_{KL}[q(o_\tau|\pi)||p(o_\tau|C)]}_{risk} + \underbrace{\mathrm{E}_{q(s_\tau|\pi)}[\mathrm{H}[p(o_\tau|s_\tau)]}_{ambiguity} - \underbrace{\mathrm{E}_{p(o_\tau|s_\tau)q(s_\tau|\pi)}[D_{KL}[q(\boldsymbol{A}|o_\tau, s_\tau, \pi)||q(\boldsymbol{A})]]}_{novelty} \end{aligned}$$

The first term, risk, is the KL divergence between the predictive posterior over observations conditioned on a specific policy and the agent's preferred observations (specified by $C$), including homeostatic set points encoded in the agents phenotype (for discussion see; Smith, Ramstead, et al., 2021; Smith et al., 2022). Smaller values therefore indicate greater similarity between the observations expected under a policy and those the agent would find most phenotypically characteristic or rewarding. The need to minimise this term promotes a preference for goal seeking behaviour. The second term, ambiguity, is the expected (conditional) entropy of the likelihood, that is, the mapping between hidden states and observations. To minimise the ambiguity term, agents will select policies that result in a precise mapping between states and observations (e.g., turning the light on in a dark room). The third and final term, novelty, is the expected KL divergence between the posterior distribution over the parameters of a model (conditioned on states, observations, and policies), and the marginal posterior over model parameters. Here, we show this for the parameters of the likelihood distribution, denoted by $\boldsymbol{A}$ (the concentration parameters of a Dirichlet prior over a categorical likelihood matrix described below), but similar terms can be included for other parameters. Here, novelty scores the shift in beliefs about the parameters of the generative model afforded by states and observations expected under each policy. Because novelty is a negative term, to minimise expected free energy, agents are driven to maximise the difference between the posterior conditioned on observations and states, and the marginal posterior, by seeking out novel observations that are expected to lead to the largest shift in posterior beliefs about model parameters. This illustrates the dual imperatives that underwrite active inference; namely, goal-seeking and information-seeking that are subsumed under a single objective function. Minimising expected free energy therefore requires agents to trade off between these imperatives; this means agents dissolve the exploration-exploitation dilemma by choosing policies that strike an optimal balance between minimising risk (i.e., maximising preferences), minimising ambiguity (i.e., maximising information gain about states), and seeking novel observations (i.e., maximising information gain about parameters); Schwartenbeck et al., 2019). For a more general formulation of expected free energy as the marginal likelihood of states under the agent's control given previous actions and observations see (Da Costa et al., 2024).

In sum, agents equipped with a limited repertoire of automatic actions (e.g., simple organisms or automated subsystems of more complex organisms) can be modelled through the minimisation of variational free energy by having an organism/agent take actions that bring about observations consistent with prior beliefs (e.g., homeostatic set-points; Buckley et al., 2017; Corcoran et al., 2020). Variational free energy can therefore be used to model perceptual inference in both continuous and discrete generative models, and simple reflex-like actions in continuous models. However, as soon as action selection requires any element of counterfactual computation, we turn to expected free energy (Corcoran et al., 2020), which (normally) requires a discrete generative model (usually a Categorical-Dirichlet model; Koudahl et al., 2021).

## 3. Generative models, belief updating, and neural dynamics

The current state of the art in active inference depicts the brain as a hierarchical "mixed model" consisting of interacting generative models (Friston et al., 2017; Parr et al., 2021, 2022; Parr & Friston, 2018c). The generative model underlying low-level sensory inference takes the form of a predictive coding network that performs inferences about continuous quantities, such as motion and contrast. Continuous low-level sensory systems then interface (via link functions that map continuous quantities to discrete latent variables) with higher-level discrete generative models (e.g., partially observable Markov decision processes, POMDPs), which perform categorical sensory inferences and select discrete action sequences (i.e., policies), which are then translated back (again through link functions) into continuous motor commands (Parr & Friston, 2018c). For tutorial-style reviews on continuous models, see Bogacz (2017) and Buckley et al, (2017). For a detailed mathematical review of discrete models see Da Costa, Parr, et al, (2020), and for a more accessible tutorial-style review, see Smith et al, (2022). For a book length treatment of both continuous and discrete models see Parr et al., (2022).

To illustrate the general principles underlying the derivation of predictive coding under active inference, here we rehearse the derivation of a single-layer predictive coding network from variational free energy assuming a static generative model and fixed priors (Bogacz, 2017; Friston, 2005). The same principles apply to the derivation of dynamical models (Buckley et al., 2017), but requires considerably more formal machinery (e.g., generalised coordinates of motion).

To arrive at a tractable expression for variational free energy we make the following three assumptions: 1) that the generative model (i.e., likelihood and prior) and approximate posterior are Gaussian; 2) that variational free energy is well approximated by a second-order Taylor series expansion around the mean of the posterior (the Laplace approximation); and 3) that variance is stationary. In fact, these are three different ways of phrasing the same assumption. Under these assumptions, the expression for (negative) variational free energy reduces to the log of the generative model evaluated at the posterior mode.

$$F = -\ln\big(p(o|s)p(s)\big) = -\ln\mathcal{N}(o; g(s), \Sigma_o) \; -\ln\mathcal{N}(s; s_p, \Sigma_p) \tag{3}$$
$$= -\ln\left[\frac{1}{\sqrt{2\pi\Sigma_o}}\exp\left(-\frac{\big(o-g(s)\big)^2}{2\Sigma_o}\right)\right]$$
$$-\ln\left[\frac{1}{\sqrt{2\pi\Sigma_p}}\exp\left(-\frac{(s-s_p)^2}{2\Sigma_p}\right)\right]$$
$$= \frac{1}{2}\left(\ln(\Sigma_o) + \frac{\big(o-g(s)\big)^2}{\Sigma_o} + \ln(\Sigma_p) + \frac{(s-s_p)^2}{\Sigma_p}\right) + C$$

Where $o$ and $s$ denote the value of sensory observations (as before) and the *a posteriori* most likely hidden state, respectively, $g(s)$ is a (typically non-linear) function mapping the value of the hidden state to sensory observations, and $\Sigma_o$ and $\Sigma_p$ denote the variance of the likelihood and prior. We next take the partial derivative of variational free energy with respect to $s$ and set up a gradient system $\dot{s} = -\frac{\partial F}{\partial s}$ equipping us with equations of motion for the value of the hidden state that moves downhill on variational free energy and in so doing approximates the true posterior over the hidden states.

$$\dot{s} = \frac{o-g(s)}{\Sigma_o}g'(s) + \frac{s_p - s}{\Sigma_p} \tag{4}$$
$$= \varepsilon_o g'(s) + \varepsilon_s$$

Line two of **eq. 4** rewrites hidden state dynamics ($\dot{s}$)—i.e., the dynamics of beliefs about the most likely hidden states—in terms of a mixture of precision-weighted prediction errors encoding the difference between the hidden state observation mapping and the hidden state prior mapping. To obtain an expression that could plausibly be computed by a neural circuit we also need two additional differential equations with neuron-like dynamics that have stationary points when $\varepsilon_o = \frac{o-g(s)}{\Sigma_p}$ and $\varepsilon_s = \frac{s_p-s}{\Sigma_p}$ leading to:

$$\dot{\varepsilon_s} = s_p \; - \; s - \Sigma_p\varepsilon_s \tag{5}$$

$$\dot{\varepsilon_o} = o - g(s) - \Sigma_o\varepsilon_o \tag{6}$$

This provides us with a set of three ordinary differential equations describing the behaviour of three neuron-like nodes whose dynamics perform a gradient decent on variational free energy. This furnishes us with a simple model of neuronal dynamics that approximate the posterior distribution over hidden states. The mapping between these equations and the cortical microcircuit is a matter of ongoing research. Broadly speaking, however, the dynamics for hidden states (implemented in "expectation nodes") are typically associated with deep layers of cortex that project laterally to error nodes in superficial layers within the same level of the cortical hierarchy and backwards to error nodes in superficial layers of subordinate levels of the cortical hierarchy. Prediction error dynamics (implemented in "error nodes") are typically associated with superficial layers of cortex and project laterally to deep layers in the same level of the cortical hierarchy and forwards to deep layers higher in that hierarchy (**Fig 1a**). The expectation node at each level serves as an observation for the level above and as a prior for the level below (for review and discussion, see Bastos et al., 2012; Hodson et al., 2023; Shipp, 2016; Walsh et al., 2020).

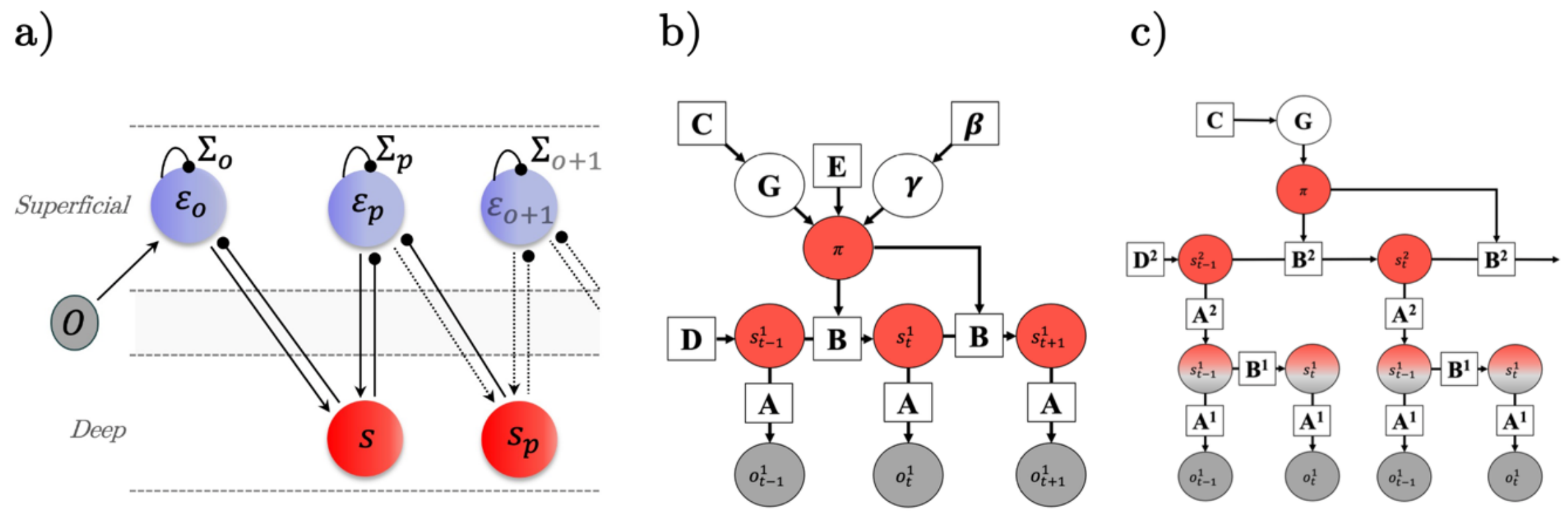

**Fig 1. Belief updating architectures. a)** Static predictive coding network, error nodes (light purple) receive excitatory feed-forward drive, and are modulated by estimates of the precision of the likelihood (i.e. the incoming evidence) and the prior. Error nodes send excitatory projections to expectations nodes (red) and receive (silencing) inhibitory feedback from the same units. **b)** Bayesian network representation of a single level, and **c)** hierarchical POMDP. Filled circles denote variables, open circles parameters, open squares functions, and arrows conditional dependence. In the hierarchical POMDP hidden states at the first level act as observations for the second level.

Moving on from continuous models, we turn next to discrete models. Unlike predictive coding networks, POMDPs (**Fig 1b**) operate in discrete time and assume discrete state and outcome spaces. Inference in these models is based upon a likelihood mapping $p(o_\tau|s_\tau)$ that quantifies the conditional probability of observations $o_\tau$ given a set of hidden states $s_\tau$, a transition probability $p(s_\tau|s_{\tau-1},\pi)$ describing how hidden states change over time dependent on the previous state and the agent's policy $\pi$, and an initial state probability $p(s_1)$. Each of these mappings is encoded in a matrix. By convention the likelihood matrix is denoted by **A**, the (prior) transition matrix by **B,** and the initial (prior) state vector by **D**. Each hidden state distribution (hidden state factor) is assumed to be independent (i.e., the mean field approximation) and assigned to a distinct **D** vector and **B** matrix. Likewise, each outcome modality (i.e. variety of sensory input) is assigned a distinct likelihood encoded in the **A** matrix. Action is modelled by making a subset of the state transitions controllable by the agent (the agent's policy space), which is why the state transitions are conditionally dependent on both the previous state and the policy. To model an agent's preferences, the generative model is equipped with a prior over preferred observations in the future $p(o_\tau|C)$, which quantifies the degree to which agents are averse to, or prefer, each observation. For each **A** matrix there is a **C** vector encoding the agent's preferences for each possible modality of observation.

As with predictive coding, the goal of perceptual inference is to infer an (approximate) posterior distribution over states given a set of observations by minimising variational free energy. Here, we use the time independent definition of (marginal) variational free energy introduced by Parr, Markovic, et al., (2019) and write the generative model in matrix form.

$$F_{\pi} = \ s_{\pi,\tau} \cdot \left(\ln s_{\pi,\tau} - \frac{1}{2}\left(\ln \mathbf{B}_{\pi,\tau-1}\, s_{\pi,\tau-1} + \ln \mathbf{B}^{\dagger}_{\pi,\tau} s_{\pi,\tau+1}\right) - \ln \mathbf{A}^{\mathrm{T}}\, o_{\tau}\right) \tag{7}$$

Again, like predictive coding, the generative model is inverted by performing a gradient descent on variational free energy with respect to states. Dropping constants, **eq. 8** gives the expression for the marginal free energy gradient.

$$\nabla_{s_{\pi,\tau}} \mathrm{F}_{\pi} = \ln s_{\pi,\tau} \ - \frac{1}{2}\left(\ln \mathbf{B}_{\pi,\tau-1}\, s_{\pi,\tau-1} + \ln \mathbf{B}^{\dagger}_{\pi,\tau} s_{\pi,\tau+1}\right) - \ln \mathbf{A}^{\mathrm{T}}\, o_{\tau} \tag{8}$$

With an expression for the free energy gradient, we can now write down an algorithm that: 1) performs a gradient decent on variational free energy with respect to states, thereby inferring the value of the approximate posterior over states; and 2) furnishes us with a simple model of neuronal dynamics (for details, see Friston et al., 2017; Parr, Markovic, et al., 2019). To this end, we define a "depolarisation" variable $v_{\pi,\tau} = \ \ln s_{\pi,\tau}$ (**eq. 9**) which represents the mean membrane potential of the neuronal population encoding the posterior distribution over states (like voltage, this variable $v_{\pi,\tau}$ takes both positive and negative values when we leave the log probability unnormalised). We then use the free energy gradient $\nabla_{s_{\pi,\tau}} F_{\pi}$ to define the dynamics of the neuronal population by setting up a difference equation that adds the free energy gradient $-\nabla_{s_{\pi,\tau}} F_{\pi} = \varepsilon_{\pi,\tau}$ (**eq. 10**) to the membrane potential $v_{\pi,\tau}$ (**eq. 11**), moving the log posterior over states in the direction of steepest decent on variational free energy. Finally, for the membrane potential to be interpretable as a probability distribution, it needs to be normalised by passing it through a softmax (normalized exponential) function (**eq. 12**). We interpret this final step as giving the normalised firing rate of the underlying population. Pulling this all together (**Fig 2a** and **eq.9** $-$**eq.12**), we have a simple iterative algorithm for inferring the approximate posterior.

$$v_{\pi,\tau} \leftarrow \ \ln s_{\pi,\tau} \tag{9}$$

$$\varepsilon_{\pi,\tau} \leftarrow \frac{1}{2}\left(\ln \mathbf{B}_{\pi,\tau-1}\, s_{\pi,\tau-1} + \ln \mathbf{B}^{\dagger}_{\pi,\tau} s_{\pi,\tau+1}\right) + \ln \mathbf{A}^{\mathrm{T}}\, o_{\tau} - v_{\pi,\tau} \tag{10}$$

$$v_{\pi,\tau} \leftarrow \ v_{\pi,\tau} + \varepsilon_{\pi,\tau} \tag{11}$$

$$s_{\pi,\tau} \leftarrow \ \sigma(v_{\pi,\tau}) \tag{12}$$

The use of the softmax function (which is a generalisation of the logistic function to vectors) to simulate average firing rate is based on the assumption made in mean-field models of large-scale brain dynamics that the average firing rate of a population can be treated as a sigmoid function of the average membrane potential (Breakspear, 2017; Da Costa et al., 2021; Hopfield, 1982; Wilson & Cowan, 1972). Event-related potentials (ERPs) in electroencephalography (EEG) research, and local field potentials in intracranial recording studies are both taken as the time derivative (i.e., rate of change) of the normalised firing rate. For a brief sketch of a possible neural implementation, see **Fig 2b.** For in depth discussion and review, see Parr and Friston, (2018a).

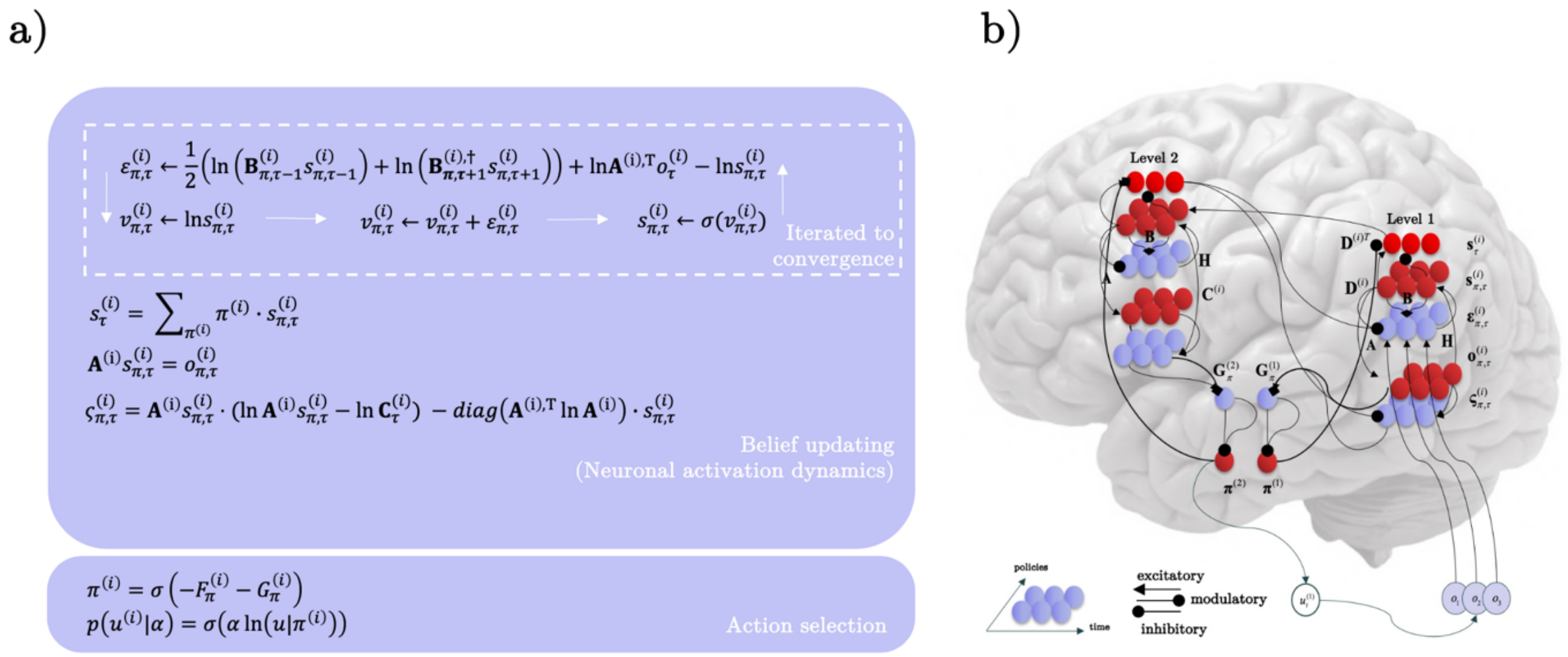

**Fig 2. Belief updating and neuronal dynamics in a POMDP. a)** Message passing and belief updating equations underlying hidden state and policy inference in a POMDP. **b)** Potential neural implementation of the belief updating scheme in two idealised cortical columns that interface with a subcortical network to perform policy selection.

So far, we have derived two simple algorithms for minimising variational free energy in cases of continuous and discrete perceptual inference, respectively: each of which have straightforward interpretations in terms of neural dynamics. However, as outlined above for action selection, we have not, by definition, received (future) observations and must therefore minimise expected free energy (**eq. 13**) to select policies. We show expected free energy below in matrix form to highlight the link between the generative model components and the components of expected free energy.

$$G_{\pi,\tau} \approx \sum_{\tau} \left( \underbrace{\mathbf{A}s_{\pi,\tau} \cdot (\ln \mathbf{A}s_{\pi,\tau} - \ln \mathbf{C}_{\tau})}_{risk} - \underbrace{diag(\mathbf{A}^{\mathrm{T}} \ln \mathbf{A}) \ \cdot s_{\pi,\tau}}_{ambiguity} - \underbrace{\mathbf{A}s_{\pi,\tau} \cdot \mathbf{W}s_{\pi,\tau}}_{novelty} \right) \tag{13}$$

Where $\mathbf{W} := \frac{1}{2}\left(\boldsymbol{a}^{\odot(-1)} - \boldsymbol{a}_{sums}^{\odot(-1)}\right)$ with $\odot$ denoting the element-wise power, $\boldsymbol{a}$ is a matrix containing the concentration parameters of the likelihood, and $\boldsymbol{a}_{sums}$ is a matrix with the same dimensions as $\boldsymbol{a}$ but where each entry is the sum of the values of the associated column in $\boldsymbol{a}$. The matrix $\boldsymbol{a}$ is updated on each trial by accumulating the co-occurrence of states and observations in a neurobiologically plausible fashion (i.e., associative plasticity). One can understand the operation of the novelty term heuristically by noticing that when agents have extensive exposure to a set of observations, corresponding to large values in $\boldsymbol{a}$ and $\boldsymbol{a}_{sums}$, the $\mathbf{W}$ term is small as it holds the inverse values of $\boldsymbol{a}$ and $\boldsymbol{a}_{sums}$, generating a small novelty value. In contrast, when agents have a limited exposure to a set of observations, the $\mathbf{W}$ term is large. As the novelty term is negative, agents are driven to maximise novelty by seeking out state-observation pairs they have had limited exposure to. For a derivation, see Da Costa, Parr, et al, (2020) and for worked examples and discussion, see Smith et al, (2022).

The posterior over policies for one timestep policies is then simply a softmax function of expected free energy (**eq. 14**), which effectively translates expected free energy into a probability distribution, from which the next action can be sampled.

$$q(\pi) = \sigma(-\sum_{1}^{T} G_{\pi,\tau}) \tag{14}$$

Here, T denotes the future time horizon of the policy. The policy that minimizes (the path integral of) expected free energy, will have the highest posterior probability. Actions are then selected by sampling from the posterior over policies at each timestep.

## 4. Computational models of conscious phenomena

Having outlined the key quantities driving the perception-action loop, the structure of continuous (predictive coding-based) and discrete (POMDP-based) generative models, and the relationship between model dynamics and measures of neural dynamics (e.g. ERPs and firing rates), we turn now to a discussion of previous work that has employed active inference to model tasks paradigmatic to the science of consciousness. In general these studies have focused on conscious contents and conscious self. For brevity, we focus exclusively on numerical (i.e., simulation) studies at the expense of the many valuable qualitative and conceptual models relevant to the neuroscience of consciousness (e.g., Ciaunica et al., 2022; Safron, 2020; Seth & Tsakiris, 2018), and accounts of the metaphysics of consciousness implied by active inference (Friston, Wiese, et al., 2020; Ramstead et al., 2023). For relevant reviews of content not covered in this article, see (Nikolova et al., 2022; Ramstead et al., 2023; Rorot, 2021). In addition, we note that although the models we review focus on visual and interoceptive sensory modalities, this selection reflects the visual and interoceptive focus of the study of conscious contents and conscious self generally. It does not reflect limitations in the applicability of the active inference framework to other sensory modalities. Indeed, to illustrate the explanatory generality of the framework — and to highlight potential applications to conscious state at the end of this section — we extend an existing model of the conscious processing of auditory regularities to account for the disruption to this process in sleep and anaesthesia.

### *4.1 Models of conscious content*

In the preceding section, we cast the perception-action loop as an iterative process of 1) inferring the approximate posterior that best minimises variational free energy (placing an upper bound on model evidence); and 2) sampling the world in a way that minimises expected free energy (thereby maximising model evidence). This motivates a perspective change on several key phenomena in consciousness science. It draws attention to the role of active sampling and anticipation in many perceptual phenomena that have commonly been thought of as largely passive.

The shift from passive to active perception is particularly prominent in models of bistable perception, which typically cast perceptual switches as noise and/or adaptation-driven oscillations (e.g. Moreno-Bote et al., 2007; Wilson, 2007), rather than states that are, at least to some extent, driven by an agent's actions. For example, Parr et al, (2019) proposed a model of two instances of Troxler fading and binocular rivalry (**Fig 3a**), that casts changes in conscious contents in terms of policy driven changes in the precision of posterior beliefs that occur as a function of saccade policies (in Troxler fading) and attention policies (in binocular rivalry). Troxler fading is a phenomenon in which percepts associated with peripherally presented stimuli fade from awareness when participants are required to maintain central fixation. Binocular rivalry occurs when incongruous stimuli are presented to each eye: instead of experiencing a superposition of stimulus percepts, participants experience discrete alternations between singular percepts with only brief periods of mixing.

Under active inference, such perceptual alternations are explained by incorporating two *prima facie* mundane observations into the task-specific generative models. The first is that an agent's estimate of the precision of state transitions (i.e., **B** matrix precision) should never be entirely certain in a capricious world (cf. Hohwy et al., 2016). The second is that the precision of sensory input depends on the location of the fovea and/or the focus of attention (modelled by making the precision of the **A** matrix conditionally dependent on saccade or attentional states). The incorporation of these two simple assumptions into a generative model results in epistemic behaviour through the ambiguity term in expected free energy. In the absence of precise sensory input, uncertainty (predictive entropy) about the perceptual content of a location of visual space accumulates, increasing the epistemic value of policies (e.g. saccade or attention policies) that will solicit informative sensory input from that part of visual space. In the case of Troxler fading, the agent is forced to maintain central fixation, precluding access to precise sensory information about the periphery. Over time, this drives the posterior over states corresponding to peripheral locations in visual space toward a uniform distribution through iterative belief updates based on imprecise state transitions and sensory input. Assuming a correspondence between the posterior over states and the content of awareness, and associating perception with a mixture of high precision states this would give rise to the fading of stimuli from awareness (we discuss the related issue of distinguishing between conscious and unconscious states within active inference in **section 5**). Similarly, discrete switches between percepts typical of binocular rivalry emerge from a generative model when agents are restricted to select between (covert) attention policies that enhance the precision (i.e., the **A** matrix mapping) of the attended stimulus at the expense of the unattended stimulus (cf. the biased competition model of attention; Desimone, 1998). The agent receives precise information about the attended stimulus, resulting in a precise posterior over states for the attended stimulus and at the same time — denied precise information about the unattended stimulus — the posterior over states for the unattended stimulus dissipates towards a uniform distribution. Crucially, as uncertainty accumulates, the unattended stimulus becomes progressively more epistemically attractive via the ambiguity reduction term in expected free energy, driving an eventual attentional policy switch. Again, assuming a correspondence between the agent's posterior over states and conscious perceptual content, this results in attention-dependent perceptual switches.

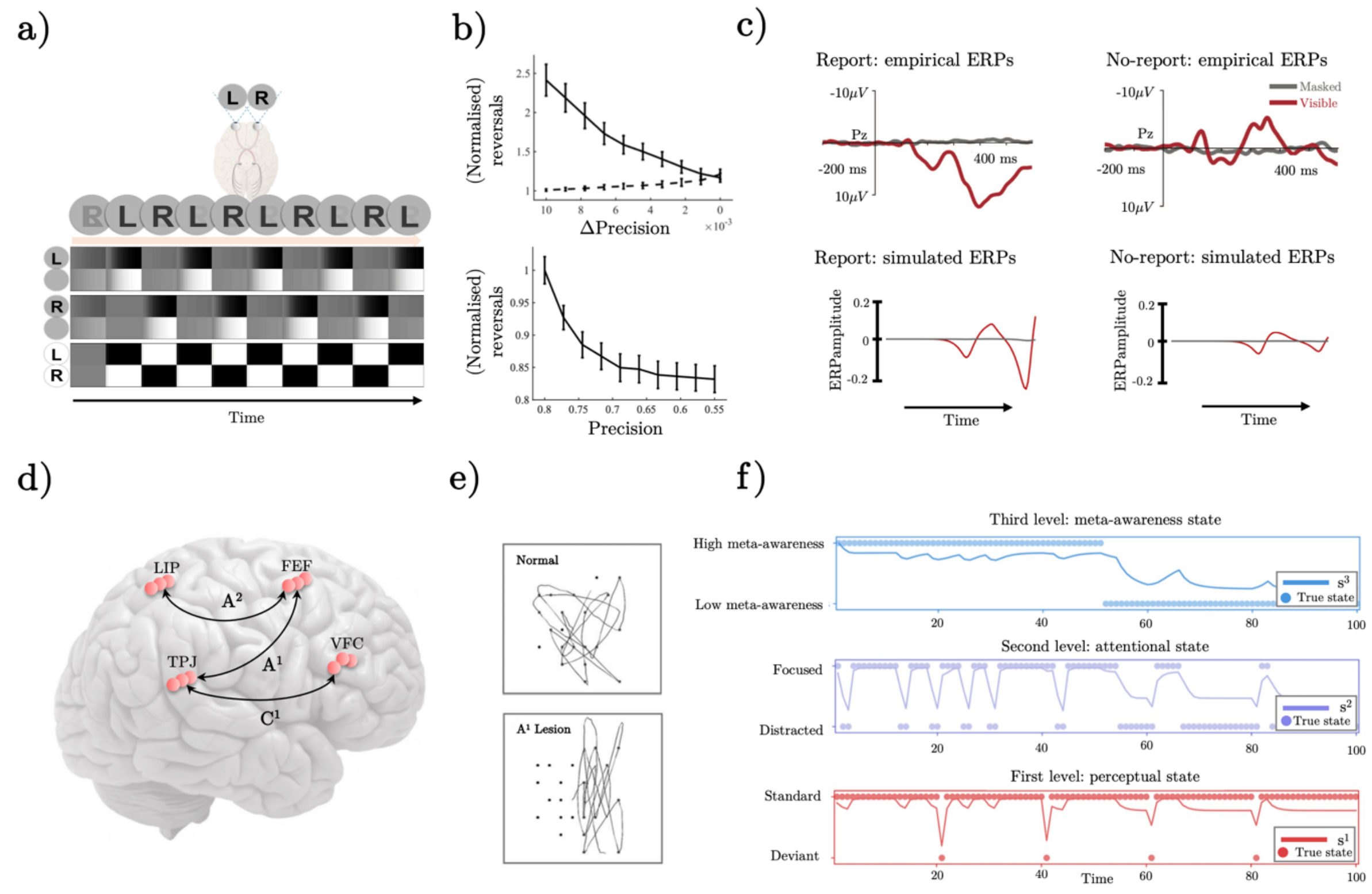

**Figure 3. Models of conscious contents and self under active inference. a)** Belief updating dynamics in a generative model of binocular rivalry. Upper rows show posterior beliefs about visual stimuli entering each eye and bottom row shows posterior beliefs about attentional state. Adapted from Parr, Corcoran et al, (2019). **b)** Simulation of Levelt's second (upper) and fourth (lower) laws under active inference. **c)** Comparison of empirical and simulated ERPs under report and no-report conditions. Adapted from Whyte et al, (2022). **d)** Macroscale connectivity scheme underlying the generative model of visual hemineglect neglect. **e)** Comparison of "healthy" and aberrant eye-movement dynamics resulting from simulated **A** matrix lesion. **d - e** Adapted from Parr & Friston, (2018b). **f)** Belief updating dynamics in a deep-parametric generative model of auditory processing adapted from Sandved-Smith et al, (2021).

This active formulation of binocular rivalry neatly explains several experimental findings that are difficult to accommodate in passive models of rivalry that do not have an explicit role for an agential process like policy selection. Specifically, the slowing of rivalry in the absence of attention (Paffen et al., 2006; Zhang et al., 2011), and the modulation of dominance durations by reward (for review see Safavi & Dayan, 2022) can both be understood as instances of expected free energy minimisation. If perceptual switches are driven by switches in attention policies, then the slowing of rivalry in the presence of a distractor task (Paffen et al., 2006) can be explained by a reduction in the precision of sensory input – thus increasing the amount of time agents need to sample the input from each eye to reach a precise posterior over states. Similarly, the addition of a reward to one of the stimuli will minimise the risk term of expected free energy in an additive manner, explaining the biasing effect of reward on rivalry (Marx & Einhauser, 2015; Wilbertz & Sterzer, 2018). In addition, although not presented in the original paper by Parr et al., (2019), in **appendix 1** and **Fig 3b** we show that the model readily explains Levelt's laws (Brascamp et al., 2015; Levelt, 1965) — a compact set of propositions that summarise the lawlike relationship between stimulus properties (e.g., luminance contrast) and perceptual dominance durations. Importantly, in addition to retrospectively explaining a large variety of existing phenomena the model provides an empirical prediction about reward driven violations to Levelt's laws (**appendix 2**) which are readily testable in the domain of human psychophysics. For an extension of this modelling strategy to the Necker cube illusion — another often-used bistability paradigm in consciousness science — see Novicky et al, (2023). For related approaches based upon reinforcement learning see (Haas, 2021; Martin et al., 2021; Safavi & Dayan, 2022, 2024). The precise relationship between reinforcement learning based models and active inference models will depend on the choice of objective functions and inference algorithms (Chou et al., 2025; Malekzadeh & Plataniotis, 2024; Tschantz, Millidge, et al., 2020). In principle, the two approaches may be formally equivalent (Da Costa et al., 2024; Da Costa, Sajid, et al., 2020). As we highlighted in the introduction, it will always be possible to derive formally equivalent, or at least very similar, models in specific domains. What active inference uniquely offers is a principled formulation of the objective function used in policy selection (i.e. expected free energy) that casts epistemic and reward-based drives within a common, information theoretic, currency that is shared across models of different tasks. Furthermore, it offers a neuronal process theory that allows one to relate belief updating processes to neuronal dynamics and plasticity.

Turning from the minimisation of ambiguity in an established generative model to novelty maximisation and parameter learning, Parr & Friston (2018b) proposed that hemineglect, a neurological syndrome characterised by a patient's neglect (i.e., unawareness) of one side of visual space (typically the left, following right hemisphere damage), can be explained by deficits in the novelty component of expected free energy. Clinically, hemineglect is often assessed through saccade cancelation tasks that require patients to circle (i.e., cancel out) all stimuli presented on a piece of paper. Here, patients with neglect will be unaware of the one side of visual space and leave stimuli on the neglected side uncancelled. To model this, Parr and Friston simulated a saccade cancelation task designed to resemble the clinical test for hemineglect. Their model used a grid to represent possible saccade locations. At the beginning of each simulation, all locations were novel (i.e., low parameter certainty), driving agents to saccade to each location and in so doing accumulate counts in the Dirichlet prior over the **A** matrix for each location (reducing the novelty component of expected free energy for saccade policies to "cancelled" locations of visual space). Lesioning the **A** matrix mapping between hidden states and visual outcomes by increasing the concentration parameters for the left side of space effectively modelled the disconnect between dorsal and ventral attention networks; removing any capacity for novelty-based saccade policies to the left side of space, thus emulating the empirical phenomenology of visual neglect (**Fig 3E-D**).

The motivation for increasing concentration parameters—or increasing confidence in the likelihood mapping—is that disconnection is a state of maximal confidence in the sense that no amount of data can change the synaptic efficacy. A key prediction of the model that is testable in healthy participants — who can update likelihood parameters via synaptic efficacy – is that individuals who become familiar with a visual scene should evoke increasingly larger belief updates, manifesting in larger event related electrophysiological responses. The underlying neural system encoding parameter certainty is expected to include afferent (feedforward) connections from the ventral attentional network in parietal cortex (TJP) to the frontal eye field (FEF) component of the dorsal attentional network. The aforementioned empirical prediction was subsequently tested in a healthy population by Parr, Mirza, et al, (2019) using a saccade cancellation task in combination with DCM and MEG. In line with model predictions, they found that when the scene was more predictable, the dorsal component of FEF actively disinhibited the (right) ventral TPJ, generating larger evoked responses. The disinhibition was, by construction of the canonical microcircuit network for DCM, most pronounced in deep layers of TPJ, which is known to receive descending input from FEF.

So far, we have discussed three general examples – Troxler fading, binocular rivalry, and visual neglect – where attention and saccade policies (arguably) play a defining role in determining the content of consciousness (For a related reinforcement learning based approach to attention see (Chalk et al., 2013). Content is determined via a process of selective sampling that generates high precision input about one location in the visual scene at the expense of other locations. However, the relationship between attention and changes in conscious content is far from one-to-one. Empirically, there is evidence that attended items can remain unconscious (Koch & Tsuchiya, 2007), and that certain stimuli can trigger corresponding perceptual experiences in the near absence of attention (Matthews et al., 2018). What then separates conscious from unconscious percepts under active inference? Plausibly, the answer comes from the hierarchical structure of the generative model underlying perceptual synthesis and the resulting separation of timescales in belief updating.

In an effort to link active inference to the broader literature on the neural correlates of consciousness, Whyte and Smith, (2021) developed a two-level POMDP model of visual awareness. This model cast conscious perception as relying on the bidirectional propagation of precise posteriors between different hierarchical levels in a generative model. Importantly, the second level of the model had sufficient temporal depth to generate goal-directed actions (such as subjective reports of the visual scene), which necessarily evolve over a longer timescale than stimulus presentation. Using this model, they simulated Dehaene et al, (2006)'s empirically derived taxonomy of the relationship between attention, stimulus strength, and conscious access. By manipulating the precision of the $\mathbf{A}$ matrix mapping between first-level hidden states and observations, representing the interplay between stimulus strength and attention, they replicated the nonlinear scaling of subjective reports and neural correlates. Consistent with empirical findings, the nonlinear increase in reported stimulus visibility was linked to a high firing rate and P3b-like ERP at the second level, resembling an "ignition" response in frontoparietal regions. Expanding on Dehaene et al, (2006)'s taxonomy, Whyte and Smith (2021) introduced expectation into the taxonomy, predicting that valid expectations would reduce P3b amplitude compared to neutral and invalid conditions, when attention is present and the stimulus well above threshold. This prediction was independently confirmed by Schlossmacher et al, (2020).

Initial work on the neural correlates of consciousness (including the work forming the backbone of Dehaene's taxonomy modelled by Whyte and Smith (2021)), supported the idea that both PFC activity and late-ERPs (e.g., the P3b) robustly correlate with conscious perception, with studies across a variety of paradigms reporting similar results (Bisenius et al., 2015; Sergent et al., 2005). However, the advent of no-report paradigms (Tsuchiya et al., 2015) challenged these findings. Under such conditions, PFC activity diminished or disappeared (Brascamp et al., 2015; Frassle et al., 2014), and late ERPs such as the P3b, once considered indices of conscious access, were no longer present (Cohen et al., 2020; Pitts et al., 2014; also see **Fig 3C**). Findings like these have led some researchers to reject PFC engagement as a necessary condition for conscious perception (Boly et al., 2017). Importantly, however, the evidential pendulum has begun to swing back in the other direction, with subsequent evidence from non-human primate electrophysiology showing that even in the absence of report, the contents of consciousness can be decoded from PFC (Kapoor et al., 2020), and that fluctuations in PFC activity precede perceptual switches (Dwarakanath et al., 2020), which is suggestive of a causal role. In an attempt to reconcile these findings, Whyte et al, (2022) modified their previous model of conscious access so that the working memory requirements of report were treated as a variety of mental action arrived at through policy selection. Available policies regarding what information to maintain in working memory controlled both the precision of the second-level **A** matrix that mapped first-level stimulus states to the second level of the model (i.e., corresponding to the gating of information from visual cortex into working memory), and the precision of the second-level **B** matrix (corresponding to the voluntary maintenance of items in working memory).

Whyte and colleagues used this modified model to simulate a visual masking task that had both report and no-report conditions. These simulations recapitulated the neural correlates of awareness under report conditions, wherein "consciously perceived" stimuli were accompanied by high firing rates at the second level of the model and a large P3b-like ERP. Crucially, Whyte et al, (2022) were able to assess the visibility of the stimulus in the model in the absence of report by simulating the task across a large range of stimulus precisions (i.e., **A** matrix precisions), thus allowing them to construct a mapping between the posterior probability at the second level of the model and the corresponding report frequency. Simulating the same task in the no-report condition, when the model did not have to provide explicit reports of its own perceptual states (and therefore did not have to maintain items in working memory), they found that the model expressed lower second-level firing rates (i.e., reduced prefrontal activity), and did not generate a P3b-like ERP, because of the reduction in message passing precision (**Fig 3C**). Importantly, however, the posterior at the second level of the model was still well above the threshold for close to 100% reportability of the stimulus, reproducing the key finding that in the absence of report, conscious access is associated with reduced prefrontal activity and no late ERPs.

This model makes two key predictions. One, prefrontal activity and late ERPs should dissociate as a function of reporting instructions, since imposing a requirement for report means that agents must increase the precision of the messages being passed between sensory and prefrontal cortices (as well as within prefrontal cortices) in a goal-directed manner; thereby altering the neural correlates of conscious access. Indeed, this is exactly what was observed in a simultaneous EEG-fMRI experiment . Namely, when the stimulus was conscious, but not task relevant, there was strong activation of visual regions, and a large N170, but only weak prefrontal activation and no P3b. In contrast, when the stimulus was conscious, and task relevant, there was strong prefrontal activation, and a large P3b. The second model prediction is that the feed-forward component of the bidirectional messages passed between prefrontal and visual cortex should contain precise information about the content of consciousness even in the absence of report, a prediction subsequently confirmed by Rowe et al, (2024).

### *4.2 Models of conscious self*

We move now from models of exteroceptive awareness, to models of meta-awareness, interoception, and emotion, where agents infer and make policy decisions about their own internal cognitive and bodily states. As consciousness is inherently subjective, self-related processing is a key area of consciousness science. It is also an area of interest to clinicians, as disturbances to interoceptive inference and policy decisions are closely tied to psychiatric symptoms and phenomenology (i.e. aberrant inferences about self-efficacy in depression (Barrett et al., 2016; Ramstead, Wiese, et al., 2023) or disrupted policy selection in rumination (Hesp et al., 2020)). For conceptual clarity, we organize the discussion thematically rather than chronologically. We direct readers interested in related approaches to (Ainley et al., 2016; Allen, 2020; Barrett & Bar, 2009; Barrett & Simmons, 2015; Critchley & Garfinkel, 2017; Theriault et al., 2021)

Based upon initial work by Seth (2013) proposing a basis for emotional content in interoceptive inference — and later work by Stephan et al, (2016) linking disorders of homeostasis and allostasis to fatigue and depression — Tschantz et al. (2022) conducted a series of simulations interrogating the capacity of different generative model structures to account for homeostatic and allostatic processes (cf. Corcoran et al., 2020). Reactive autonomic responses to changes in bodily state were modelled by a predictive coding network, which were generalised to simple forms of anticipatory action by conditioning the homeostatic setpoint on the inferred exteroceptive state. Crucially, alterations in the precision of interoceptive prediction error affected sensitivity to interoceptive state changes, leading to failures in homeostatic adjustment. The balance between descending proprioceptive prediction errors and ascending interoceptive prediction errors dictated whether the agent adjusted its prior beliefs about physiological state or executed autonomic actions to align physiological states with set points. Favouring ascending sensory prediction errors over autonomic adjustments—i.e., when the prior over the current set point is sufficiently imprecise compared to the anticipated interoceptive data—prompted agents to change their beliefs about the set points themselves. This failure of autonomic regulation may speak to paradoxical sensory dysfunction in conditions like autism (Gu & FitzGerald, 2014), where hypersensitivity to sensory input co-occurs with diminished autonomic responses and aberrant allostasis. Finally, employing a POMDP to interface with a predictive coding network allowed the model to account for goal-directed interoceptive control, where agents make decisions about actions by anticipating deviations between the preference for maintaining zero body temperature (the preferred homeostatic setpoint) and the anticipated body temperature.

Complementing the insights from this proof-of-principle modelling approach, recent studies fitting active inference models to empirical data have inferred disrupted interoceptive precision across various disorders, including depression, anxiety, eating disorders, and substance abuse disorders (Smith et al., 2020). Notably, in a heartbeat tapping task, the best-fitting model (a simple hidden Markov model inverted through the minimization of variational free energy) exhibited a failure to adjust interoceptive precision in the face of a breath-hold perturbation in the patient sample, whereas the healthy controls successfully increased their interoceptive precision. This finding was recently replicated in a preregistered study with a large transdiagnostic sample (Lavalley et al., 2024) and in a cohort of healthy controls (Smith, Kuplicki, Teed, et al., 2020).

In an important study testing some basic predictions of the neural process theory associated with active inference, Smith et al, (2021) leveraged a novel gastrointestinal perception paradigm that had participants report the presence (or absence) of vibrations of varying magnitude in their stomachs delivered via an ingestible vibrating capsule whilst recording neural responses with simultaneous EEG. Inspection of the best-fitting model parameters (again a simple hidden Markov model) found evidence in support of the neural process theory. Specifically, despite not being fit to either reaction times or neuronal responses, increased interoceptive precision (**A**-matrix precision) was positively correlated with participant reaction times and the magnitude of evoked responses in sensory (parietal-occipital) electrodes.

Shifting to a more cognitive context, Allen et al, (2022) constructed a model of cardio-visual sensory integration that inferred the phase of its cardiac cycle (diastole vs. systole) when presented with arousing or non-arousing visual stimuli. Based on stimulus type, the model inferred its cardiac policy, controlling the state transitions between cardiac cycles. The hidden state of the cardiac cycle, in turn, controlled the precision of the visual **A** matrix. This minimal model reproduced several otherwise unrelated empirical findings. Arousing stimuli led to immediate cardiac acceleration (defensive startle reflex; Graham & Clifton, 1966) and a synthetic lesion to the **A**-matrix mapping, corresponding to a reduction in interoceptive precision, produced "psychosomatic hallucinations" and increased false inferences or metacognitive biases (e.g., Allen et al., 2016; Hauser et al., 2017). In support of the assumption underlying the model — that agents infer their cardiac policy based upon the visual stimulus they are presented with — Corcoran et al., (2021) found that both tonic heart rate and (high frequency) heart rate variability decreased as a function of sensory ambiguity.

Formalising a body of theoretical and empirical work relating interoceptive, perceptual, and contextual inferences to emotion, Smith et al, (2019) showed that active inference agents could acquire a range of emotion concepts over the course of an "in silico childhood". In particular, each agent started with a flat **A** matrix mapping between hidden emotion states ("emotion concepts") and interoceptive observations, and over the course of hundreds of trials consisting of observed conjunctions of interoceptive observations of arousal, valence, and behaviour, and exteroceptive observations of context, the agent learned a mapping between hidden emotion states and interoceptive observations. Interestingly, they found that an impoverished "in silico childhood", in which the outcome statistics that the model was exposed to were biased to one particular emotion (e.g., sadness), led to reduced accuracy of the model in a subsequent emotional inference task, even after being exposed to other emotions. In a similar but more empirically focused context, Smith, Lane, et al. (2019) used a hierarchical model to simulate inference in an emotional working memory task. The agent was required to categorise, and then compare, two successive emotional states. Intriguingly, they showed that even in their relatively simple model there were at least seven distinct underlying neural mechanisms capable of producing the phenotype of reduced emotional awareness. For example, when the agent had a high prior expectation of uncharacteristic bodily states (as is the case in some forms of anxiety), it reliably miscategorised its own internal states of sadness and panic as sickness and heart attack, respectively.

In both of the models described above, valence was treated as an observation, not something that was itself inferred — a reasonable simplification given that the target of explanation was emotional inference (as opposed to affective inference). To complement this treatment, Hesp et al, (2021) introduced a model of affective inference proposing that valence was inferred (in part) from the mismatch between their prior and posterior over policies. The prior over policies was based on expected free energy alone (e.g. **eq.14**), while the posterior was based on expected free energy and a post-hoc belief based upon successive observations. The change from prior to posterior therefore reflects a prediction error indicating how consistent a new observation was with prior expectations under each policy. The direction of this error was then used as an observation in a higher-level model that inferred valence states. Here, lower-level observations supporting priors over policies promoted positive valence, while those inconsistent with priors promoted negative valence. This kind of (reward) prediction error or "affective charge" had been previously associated with phasic dopamine discharges (Friston et al., 2014; Schwartenbeck et al., 2015).

Simultaneously, this affective charge was used to iteratively update a precision parameter on expected free energy, for which valence states acted as priors. Here, negative valence states and unexpected observations both reduced this precision, decreasing the subsequent influence of expected free energy on policy selection. This decrease plays a few complementary roles. First, it promotes probability matching behaviour, which can be adaptive if one has low confidence in beliefs about policies or plans. Second, it permits a stronger influence of learned habits (usually denoted by $\mathbf{E}$), if specified. Finally, it optimises the relative influence of expected free energy on policy selection. Namely, when expected model evidence is higher under the posterior over policies than under the prior over policies, affective charge (and valence) is positive, indicating that the agent has an increased confidence in their plans, increasing the precision of expected free energy relative to other terms. Conversely, when expected model evidence is higher under the prior over policies than under the posterior over policies, the valence down-weights expected free energy (i.e., it reduces the contribution of risk, ambiguity, and novelty to policy selection). Crucially, because the two-level model allowed valence states to act as prior beliefs about policy precision, the agent could contextualise the contribution of expected free energy to policy selection, which improved performance in a reversal learning task. Put another way, by giving the agent a rolling estimate of the degree of confidence they should have in their internal model of the mapping between actions and the observations they generate, agents could optimally weight the contribution of current observations (variational free energy) and expected future observations (expected free energy) to the selection of policies.

The form of deep parametric inference in Hesp et al.'s work was subsequently extended by Sandved-Smith et al, (2021) to model a form of meta-awareness underlying the cyclical phenomenology of focus → distraction → awareness of distraction → focus, that is commonly studied in contemplative neuroscience. The key contribution of the model was to cast attention, the control of attention, and awareness of attentional control in terms of the hierarchical control of **A** matrix precision. **A** matrix precision at the first level of the model, mapping sensory observations to hidden states, was controlled by the attentional state of the agent at the second level, which consisted of two idealised states: 'focused' and 'distracted'. The agent had a preference (corresponding to a task instruction or goal state) for maintaining itself in a focused attention state that, through an imprecise second level **B** matrix, occasionally transitioned the agent into an unfocused state. Policy selection at the second level — mental action — allowed the agent to reorient their attentional state and transition back into a focused state, once the agent became aware of itself occupying the unfocused state: a key aspect of many forms of contemplative practice. Meta-awareness then consisted of a third level which, in turn, controlled the precision of the second-level awareness states.

Hidden states at this third level consisted of two states: high meta-awareness, which entailed a high-precision second-level **A** matrix, and low meta-awareness, which entailed a low-precision second-level **A** matrix (via a second-to-third level **A** matrix mapping). As a proof of principle, they simulated a simple auditory oddball task under differing third-level attentional states and showed that the agent spent longer in a distracted attentional state under low meta-awareness conditions than under high meta-awareness conditions. Indeed, in line with the finding that mind wandering increases under conditions of low perceptual demand (Lin et al., 2016), in the low meta-awareness state the agent only noticed the switch from focused to distracted when auditory oddballs occurred. In other words, oddballs induced larger prediction errors — and resulted in stronger ascending evidence — forcing large belief updates to the approximate posterior over hidden states across the hierarchy. Importantly, unlike the models discussed above, which aim to unify existing findings and generate novel predictions, the explanatory aim of this model is best thought of as providing conceptual clarity to areas of research, in early phases of development, where theory is largely verbally defined and has at best a loose connection to empirical data. This type of modelling is, therefore, best thought of as a kind of computational conceptual analysis that is a pre-requisite to modelling aimed at unifying existing results or providing empirical predictions.

### *4.3 Models of conscious state*

Active inference models have so far not been applied to experiments that manipulate conscious state (but see Hobson and Friston (2012) for a theoretical review). Importantly, however, manipulations of conscious state are characterised by sensory and motor disconnects (Cirelli & Tononi, 2023) that are at least *prima facie* well suited to explanations cast in the language of active inference. For example, neuronal responses to stimulus pattern violations that occur over long time scales vary across sleep and anaesthetic states while pattern violations that occur over shorter time scales are preserved (Boly et al., 2011; Dehaene et al., 2011). To highlight the explanatory generality of active inference, in this section, we present a simple extension of the hierarchical model of auditory regularity perception presented in Smith et al, (2022). We show how the targeted manipulation of precision reproduces the finding that the late P3b ERP component, typically associated with the detection of violations of "global" (long-timescale) auditory patterns in wakefulness (Bekinschtein et al., 2009), is absent across NREM and REM sleep states, whilst ERP responses to "local" violations are preserved (Strauss et al., 2015). We provide a brief overview of the model structure to contextualise the results, and refer interested readers to Smith et al, (2022) for a step-by-step description of the model.

The model consists of two hierarchical levels (**Fig 4a**); the first level tracked moment-to-moment changes in auditory tone and the second level tracked the pattern of tones at the first level, inferring the overall trial type (i.e., whether all stimuli had the same "standard" tone or whether there was an "oddball"). We allowed the model to accumulate concentration parameters in the second-level **D** vector over the course of 10 trials. On trial 10, the tone either conformed to — or violated — the expected trial type. In line with empirical findings, violations of "global" patterns induced large P3b like ERPs at the second level of the model, and violations of "local" within-trial expectations induced a mismatch negativity (MMN)-like ERP at the first level of the model. Based on the finding that NREM and REM sleep are low adrenergic states (Cirelli & Tononi, 2023), we leveraged the process theory associated with active inference to perform a targeted manipulation of the model designed to reproduce the absence of noradrenaline by reducing the precision of the second-level **B** matrix (by applying a softmax function with precision parameter $\boldsymbol{\omega} = 0.9$), which has been linked both theoretically (Parr & Friston, 2017a, 2018a), and empirically (Vincent et al., 2019), to adrenergic tone. In line with empirical findings (Strauss et al., 2015), the reduction of **B** matrix precision eliminated P3b-like ERPs at the second level of the model but left the MMN-like ERP at the first level intact (**Fig 4b**). As dynamics at the first level do not depend upon time-step by time-step dynamics, first level **B** matrix lesions have no effect on ERPs at either level of the model.

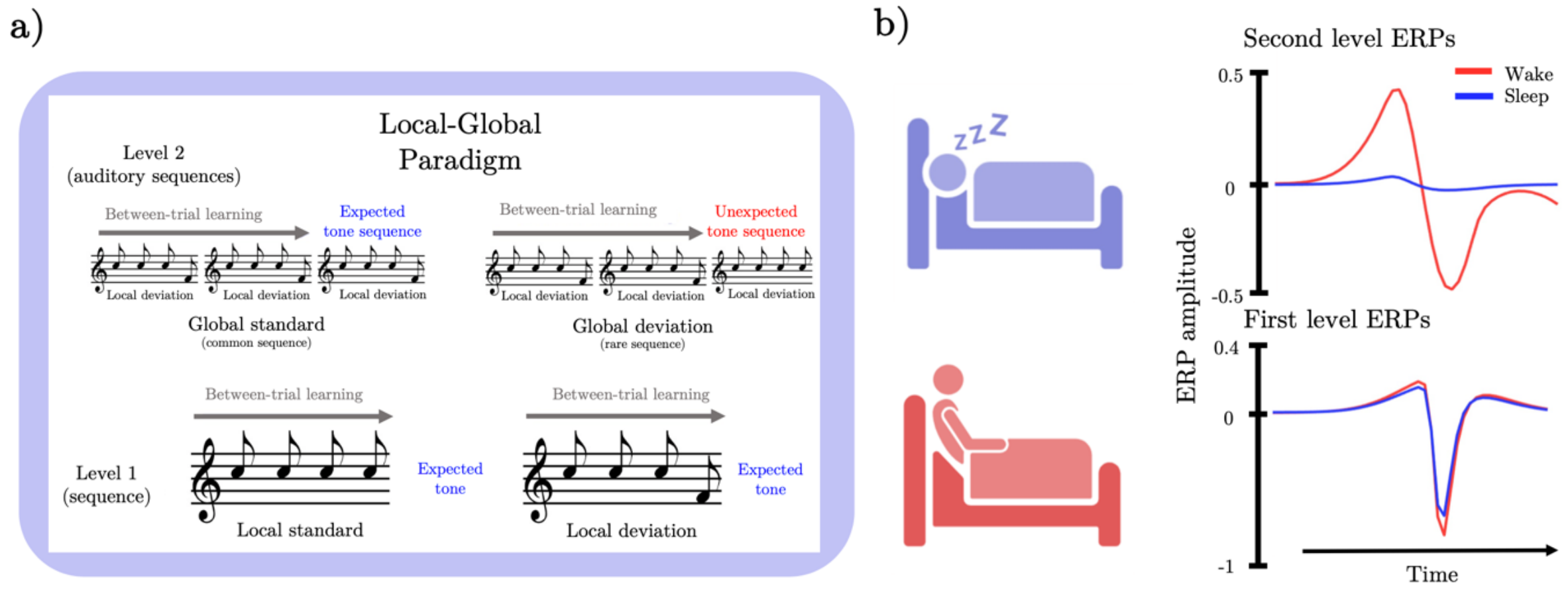

**Figure 4. Active inference model of disrupted auditory processing across sleep states. a)** Representation of the local-global paradigm encoded in the generative model of auditory regularity processing. The first level of the model infers the mapping between the tone observations and hidden states, while the second level of the model infers the longer-timescale pattern in states at the first level. The model learns the initial state probability encoded in the **D** vector across trials. Adapted from Smith et al, (2022). **b)** Simulated ERP responses to auditory pattern violations in "sleep" and "wake" conditions. First-level ERPs are relatively unaffected by the reduction in second level **B** matrix precision

(chosen to represent the low-adrenergic conditions of both REM and NREM sleep), while second-level P3b-like ERPs are close to abolished.

As described, this model is simply an explanation of the absence of late ERPs in sleep states, and does not in-itself explain why participants lose consciousness in parts of sleep. Although, if we follow the line of argument put forward in the proceeding section — that conscious access requires a level of temporal depth that abstracts away from the moment-to-moment flux in sensory input (Friston, 2018; Friston et al., 2012; Whyte & Smith, 2021) — the model on offer here may also provide an algorithmic-level explanation for why NREM sleep is associated with fewer dream reports indicative of an absence of consciousness (we discuss what makes states conscious rather than unconscious under active inference in more detail in **section 5**), and why REM sleep is characterised by more frequent dream reports with bizarre and seemingly incoherent phenomenology (Hobson, 2009).

Specifically, NREM sleep is a low cholinergic state, which under active inference would result in a disconnection between the hierarchical levels of the generative model via a reduction in A matrix precision (i.e., the likelihood mappings between successive hierarchical levels. This would effectively take the temporally deep level 'offline' and greatly minimise belief updating, plausibly corresponding to dramatic reduction or loss of consciousness in NREM sleep. REM sleep, by contrast, is a high cholinergic state which would allow hierarchical message passing to occur between levels. Crucially, a model in a high cholinergic state — with imprecise sensory constraints — combined with low noradrenergic tone would show imprecise transitions between hidden states (via low B matrix precision), plausibly explaining the bizarre phenomenology of dreams in REM sleep. Interestingly, REM sleep behaviour disorder—often thought of as a precursor to Parkinson's disease—manifests as acting out of such dreams, possibly reflecting a failure to balance precision across hierarchical levels, and particularly to suppress the precision associated with proprioceptive prediction errors during dreaming (Peever et al., 2014). Please see (Parr & Friston, 2017a) for a brief review of modulatory neurotransmitters and the precision of belief updating in the brain.

This minimal example neatly illustrates the typical strategy of modelling under active inference: start with a target phenomenon that is sufficiently well-defined to allow the construction of a generative model and, based on a combination of empirical and theoretical neurobiology, simulate a targeted manipulation of the model designed to mimic an experimental manipulation. Given the rich neurobiology of sleep and anaesthetics, this strategy can, as illustrated above, be straightforwardly applied to experimental manipulations of conscious state. The model presented here is, of course, just one example of a generative model of neural dynamics across conscious states. Ideally, future work will build and compare multiple competing models to empirical data.

## 5. Active inference as a minimal theory of consciousness

We started this paper by introducing active inference with the aspiration of building toward a theory of consciousness, using active inference models of different facets of consciousness as building blocks. With a clear idea of the framework and the type of explanations offered by active inference models in place, we now address this aspiration. To make the structure of the theory and its relation to empirical data as clear as possible, we follow recent work in the philosophy of consciousness science (Negro, 2024) and leverage Lakatos' account of 'scientific research programs' (Lakatos, 1968), which is generally viewed among contemporary philosophers of science (Godfrey-Smith, 2003) as the successor to Popper's falsificationist program (Popper, 2008).

According to Lakatos (1968), research programs have two components: a hard core and a protective belt. The hard core is constituted by the set of fundamental concepts that form the foundation of the theory and need not themselves be directly testable. The protective belt, in contrast, is made up of the less central concepts and assumptions that generate testable predictions when combined with the hard core. Changes must only be made to the protective belt of a research program if it is to survive. Importantly, the changes that are made must be progressive. That is, the changes must make the theory more precise or expand its explanatory reach to increase its predictive power. If, however, changes to a theory only seek to explain away contradictory results and do not generate new predictions or contribute other explanatory virtues, then the research program is said to be 'degenerative' and may eventually need to be abandoned. For example, the hard core of integrated information theory (Albantakis et al., 2023; Oizumi et al., 2014; Tononi et al., 2016) is constituted by the postulates that lead to the derivation of Phi as a measure of consciousness. The periphery is constituted by the bridging assumptions and approximations that make Phi computable. It would be a serious challenge to integrated information theory if Phi did not reliably co-vary with conscious state. Importantly, however, it would not be a deathblow to integrated information theory, as it may very well be that one of the assumptions underlying the derivation of a computable approximation to Phi is responsible for the failure, rather than it being a failure of the hard core of the theory. The health of the research program would then be decided by the ability of integrated information theory to alter the protective belt in a manner that generated additional predictions.

At the centre of the active inference framework is the imperative to minimise two objective functions, expected free energy and variational free energy. These two objective functions, and the imperative to extremise them, must therefore be included in the hard core of any theory that is based upon active inference. This entails that all behaviour, both conscious and unconscious, must, in some minimal sense, be aimed at minimising these quantities (cf. Hohwy, 2021). The neuronal and computational process by which this occurs is, at least currently, a part of the protective belt of the theory.

The next component we need to move towards a theory of consciousness, is a link between the quantities that enter into the objective functions and conscious experiences. The most minimal such link implicit in all the models reviewed above is between the contents of consciousness and the inferred hidden state (i.e., the approximate posterior over states $q(s)$) of the world, body, and brain. It follows therefore that all changes in conscious contents must result from a change in the inferred state of the world, body, or brain; this would include changes from not having conscious content to having conscious content, as in the model of conscious state, as well as exteroceptive and interoceptive contents, as in the models of perception and emotion and affect, and contents arising from the state of the brain itself, as in the case of meta-awareness. This is a minimal link because it only suggests that the kind of selective sampling characteristic of active inference is necessary, rather than sufficient, for a change in conscious content. The minimal link rules out that a change in consciousness could occur without a change in the posterior over states (but not that a change in posterior could occur without a change in conscious experience), and is consistent with non-conscious creatures exhibiting behaviour governed by active inference.

Finally, in order for active inference to be a theory of consciousness as such, and not just a theory of adaptive behaviour, we need a criterion to distinguish between conscious states and unconscious states (Doerig et al., 2021; Seth & Bayne, 2022). In the language of active inference, we need an account of what makes some posterior beliefs conscious and others unconscious. Here, again we look to the assumptions made in computational models of conscious content derived under active inference. In all the models reviewed above, the posterior beliefs that best correspond to the contents of consciousness are those that drive policy selection, specifically those that drive the selection of subjective report policies. Minimally, the computations underlying subjective reports require a discrete generative model capable of performing inference over mutually exclusive, distinct, states of the world. These discrete alternative states can also generate trajectories through a continuous space within mixed models (Friston et al., 2017). Continuous generative models, lacking this counterfactual (discrete) element, can drive reflex-like actions, but the types of computations underlying subjective reports (e.g. confidence wagering, or judgements of presence vs. absence) are explicitly counterfactual. As such, they require a discrete generative model capable of considering one or more mutually exclusive alternatives. Importantly, deep levels of discrete generative models evolve at too slow a timescale to match perceptual phenomenology (Whyte & Smith, 2021). We posit, therefore, that the conscious/unconscious distinction is tied to the mechanism(s) of conscious access, and that it occurs at the discrete-continuous interface where continuous posterior beliefs driven by moment-by-moment changes in the sensorium are transformed into a discrete format that can drive action selection. For a posterior belief to become conscious, it must, therefore occur at a timescale that is sufficiently slow so as to abstract away from the immediate sensory flux making it available to inform counterfactual policy selection, and also be suitably precise to reliably distinguish between different counterfactual states of the world. This sits well with previous philosophical on work active inference arguing that the computational role of conscious access in a hierarchical generative model is to force an agent to transiently settle on a particular posterior belief about the hidden state of the world in order to drive action selection (Hohwy, 2013; Marchi & Hohwy, 2022; Whyte, 2019). The precision requirement can be seen as a computational analogue of the observation that conscious experience is composed of differentiated, often mutually exclusive, perceptual objects (Canales-Johnson et al., 2017; Oizumi et al., 2014; Seth, 2021; Seth, 2014; Tononi, 1998). Crucially, this also implies that there is no hard threshold at which point a posterior belief will become available for report. Rather, the threshold for conscious access will vary with task demands and the type of decision an agent is required to perform. In addition, we wish to be clear that we are not claiming that being the target of a counterfactual policy selection process is what makes a posterior belief conscious. Rather it is the capacity (in a counterfactual sense) for such a posterior belief to inform policy selection processes at the right timescale and with sufficient precision that makes it conscious (c.f. dispositional theories of consciousness; Carruthers, 2003; Prinz, 2012). We regard this as a necessary and sufficient condition for content to become conscious in humans (and non-human animals with similar neuronal architectures) but we remain agnostic as to whether this condition is both necessary and sufficient in systems vastly different to ourselves. The specifics of the decision criterion that best explains subjective reports is a matter of ongoing empirical debate and is best thought of as a part of the protective belt of the theory (for a possible implementation see Whyte et al., 2022). For a discussion of the relationship between phenomenology and the depth of counterfactual policy selection see (Seth, 2014).

Importantly, the minimal links between conscious processes and active inference contained in the hard core of the theory should be thought of as foundational building blocks in the discovery and construction of a more comprehensive theory of consciousness currently implicit in the periphery of active inference. The first two components of the theory (the imperative to minimise variational free energy and expected free energy, and the link between conscious perception and the posterior over states) are in some sense logically entailed by the formal machinery of the active inference framework. The third component of the hard core (that conscious access occurs at the discrete-continuous interface) is, in contrast, an inference to the best explanation based upon the types of generative model needed to model paradigmatic experimental paradigms used to study consciousness. In other words, although there are hints in the fundamental setup of active inference, a complete theory of consciousness needs more than the principles of optimising an internal model through action and perception. It must specify the properties of the model itself to be scientifically useful. The process of discovery involves incorporating, testing, and discarding models (including those reviewed above) and including further refinements to existing models (such as the modifications made to the model of binocular rivalry and hierarchical auditory processing; see **appendix 1-2**). Inspection of these models then informs a richer theory, suggesting which properties are pivotal in explanations of conscious content, self, and state.

The approach we advocate here is first and foremost one of open-ended theory-building and discovery, where the theoretical construct is continually expanded and finessed, as long as it remains a progressive research program, and stays in the race for model evidence against other theories of consciousness (see Corcoran et al., 2023). Key future developments will include modelling further conscious phenomena in more detail, including the application of structurally similar models to unify diverse conscious and unconscious phenomena. This is exemplified by the model of rivalry and Troxler fading (Parr, Corcoran, et al., 2019); and by the hierarchical expansion of the model of attention-driven perception used to accommodate no-report paradigms (Whyte et al., 2022). Further work should also investigate the performance of models that combine elements from models of conscious content, self, and state. Additionally, it will be important to refine conceptual models that target distinctive kinds of phenomenology (e.g., emotion and affect, objecthood, meta cognition; Barrett et al., 2016; Nikolova et al., 2022; Pezzulo et al., 2018; Seth, 2014; Seth et al., 2012; Seth & Friston, 2016) to allow greater computational specificity and the ability to generate testable predictions.

The protective belt of the theory is constituted by the formalisms that allow specific models of conscious phenomena to be written down and simulated, and the assumptions that map the results of the simulations to empirically derived behavioural and neural data. This includes the structure of the generative model, the process theory that maps the generative model and the dynamics of state and policy inference to neural data, and the assumptions that link the quantities arrived at through simulation to reported phenomenology (e.g., does the content of perception correspond to the mode of the posterior and/or does the precision of the distribution contribute to sensory phenomenology?).

This current framing of the theory is deliberately minimal, leaving room for substantive disagreement between active inference theorists about issues that are empirically or theoretically underdetermined. It also acknowledges that the vast majority of explanatory work will be done by the protective belt of the theory and, as such, is largely open for productive revision and refinement (given the repertoire of possible generative model structures as well as the flexibility of the process theory mapping between simulated and empirical data). At the level of individual generative models, it is almost always possible to formulate a large set of candidate models that can explain the relevant empirical data to a greater or lesser degree, and in extreme cases may actually generate mutually contradictory predictions. This is a feature, not a bug, and is complemented by the fact that active inference as a modelling framework has been developed alongside a suite of variational (Bayesian) methods for model fitting and model comparison (Daunizeau, 2011; Friston et al., 2007; Zeidman et al., 2022). Such methods make it possible to adjudicate between families of competing generative models, each of which embody a hypothesis about the processes generating observed empirical data. The development and deployment of individual generative models — and the development of active inference as a theory of consciousness — therefore go hand in hand, leaving room for the theory to both contribute to, and evolve alongside, empirical debates.

For example, as we alluded to above, the variety of decision criteria underlying conscious access to perceptual states is a matter of ongoing discussion, with recent evidence pointing to a potential metacognitive mechanism involving abstract inferences about the presence or absence of a stimulus independent of the structure of the perceptual space itself (Dijkstra et al., 2023). This mechanism is contrary to a previously proposed active inference architecture (Whyte et al., 2022), which modelled conscious access through a policy space that is explicitly linked to the structure of the agent's perceptual space. Current evidence is insufficient to disambiguate between the two hypotheses in a more general setting, but if conscious access does indeed turn out to rest on a kind of metacognitive decision independent of the structure of a perceptual state space, then this would pose a significant (possibly insurmountable) challenge to this specific active inference model. Crucially, however, this component of the model is idiosyncratic and is not an intrinsic feature of the structure of generative models under active inference, making it possible to build future models that respect novel findings (and in so doing, productively extend the explanatory reach of active inference into the metacognitive domain). We regard this flexibility as an explanatory virtue of the theory rather than a deficit, and one that is entirely appropriate given the early stage of the science of consciousness and the ongoing revision of empirical findings as methods and data quality improve.

These considerations leave open the worry that active inference may be too flexible to be productive as a theory of consciousness (rather than just a useful modelling framework). Or, put another way, the link between the hard core of the theory and its protective belt leaves too many degrees of freedom to be able to lead toward a recognisable theory of consciousness in its own right. There are, however, two key considerations which assuage this worry. First, the aforementioned example of a potential revision to an active inference model of conscious access involves revising an assumption about an agent's policy space in only one model, and thus does not represent a substantive blow to the protective belt of the theory. Crucially this is not true of all assumptions. The fact that distinct models share objective functions which are composed of a small number of semantically interpretable quantities (i.e., accuracy, complexity, risk, ambiguity, novelty) limits the number of parameters that can plausibly contribute to specific behaviours. For example, in modelling selective attention, the salience of a stimulus has been attributed to the precision of the **A** matrix – which will exert an influence over behaviour through the epistemic (ambiguity minimising) component of expected free energy. This assumption is shared by almost all active inference models that have a role for attention and visual search (Allen et al., 2019; Holmes et al., 2021; Mirza et al., 2018, 2018, 2019; Parr et al., 2021; Parr & Friston, 2017a, 2017b, 2018b; Whyte et al., 2022; Whyte & Smith, 2021). Finding an example of an attentional effect that cannot be modelled in this way would therefore represent a very substantial blow to the active inference theory of consciousness, as the effect of the revision would impact the vast majority of existing models.

The second important consideration is that there are, in fact, both quantitative and qualitative predictions generated by the theory. Although not derived directly from the hard core, these predictions are close enough that if they were found to be false, they would require the revision of bridging assumptions made in the vast majority modelling studies – even those that may appear to have nothing to do with consciousness science. A revision of this kind would put the theory dangerously close to being a degenerate research program (Godfrey-Smith, 2003). The existence of an explore-exploit trade off in active vision is one quantitative example of such a prediction. We give an explicit example of such a prediction in the context of binocular rivalry in (**appendix 2**). If it were discovered that the reward associated with a stimulus did not trade off against the epistemic content of a competing stimulus (e.g., luminance contrast) in the selection of attention policies then active inference would be seriously challenged, as the trade off between risk and ambiguity in expected free energy is a fundamental component of the hard core of the theory. Of course, this prediction would be difficult to test outright, since there are individual differences between participants in learning rates, pessimistic priors, and susceptibility to reward association, among other things. Also, the extent to which attention plays a role in early visual processing is still uncertain, meaning that, in practice, it will often be necessary to make an inference to the best explanation across a wide variety of experiments. If, however, it was robustly shown that the best fitting models across participants and experimental paradigms (e.g., binocular rivalry and Troxler fading) did not respect the expected trade off between risk and ambiguity when selecting attention policies, then the active inference theory of consciousness would become less plausible. Hence, there is a sense in which the active inference modelling framework —and the model fitting and comparison tools that accompany it – hold the key to serious testing of active inference as a theory of consciousness. This is a virtuous position to be in. The tools for model fitting and comparison in no way assume the truth of the theory used to derive the models being fit. The best fitting model in a specific context could very well be a model that violates core tenets of the active inference theory of consciousness.

We can also derive a second qualitative prediction from the theory. Specifically, we noted above a core commitment of active inference as a theory of consciousness is that if there is any change in conscious contents, then there must be a change in the inferred state of the body, brain, or world. A corollary of this commitment is that, in the absence of sensory (**A** matrix) precision, posterior beliefs about the state of the world should not be moved from their prior trajectories (e.g., our ability to silently count in our heads implies we can change our posterior beliefs moment to moment based upon priors about the transitions from one number to the next, but we would not expect to deviate from this numerical trajectory in the absence of precise sensory data), and therefore, conscious contents should not change. A handy slogan for the theory is therefore that “To See is to Look” (and likewise for other sensory modalities, such as “To Hear is to Listen”, “To Feel is to Touch” etc.). That is, to consciously perceive is to employ precision assignments characteristic of some variety of counterfactual policy selection process, across overt and covert actions. The complete assignment of zero precision to a sensory modality is likely too much of an idealisation to be empirically tractable; however, a related prediction that is nonetheless related to the core of the theory is that a reduction in precision, will attenuate belief updating; manifesting as delayed changes in conscious contents. If we associate the endogenous assignment of precision to covert and overt attention policies (Hohwy, 2012), which is an assumption that is made throughout active inference models of active vision, then a stimulus that has been assigned high precision in the selection of an attention/saccade policy will enter the contents of consciousness sooner than a retinally-matched stimulus that is not the target of a saccade. This prediction is also unique to active inference as a theory of consciousness and stands in contrast to other potentially more targeted theories of consciousness such as global workspace theory or integrated information theory, which have no explicit role for action, whereas under active inference the allocation of precision comes through the explicit selection of attention or saccade policies. This kind of empirical prediction is indeed part of an adversarial collaboration testing integrated information theory against both passive predictive processing theories of conscious content and active inference (INTREPID CONSORTIUM, 2021; **project number TWCF0646**).

Finally, it is important to note that the success or failure of the theory does not depend entirely on the success of these two close-to-core predictions. It would be possible for these predictions to be validated and active inference to still fail as a theory of consciousness. Indeed, there are a variety of less dramatic — and, we suspect, more likely — ways the theory could fail. For instance, it would be indicative of a degenerative research program if active inference models of key phenomena were overly complex (in the sense that they overfit to idiosyncratic experimental results and do not generalise), or if models of conscious and non-conscious perception look indistinguishable, or if models of different conscious phenomena fail to display consistent commonalities and differences, such that the theory would offer little in terms of a unifying or distinctive explanation. In turn, if the theory should turn out successful, we would expect to eventually be able to reproduce all the key behavioural and neural correlates of consciousness with a relatively small set of models that vary along a small number of dimensions reflecting differences and similarities in reported phenomenology and neural correlates. This also implies a direction of explanatory travel in the nature of the explanation. Neural and behavioural variables are mapped to phenomenology (and vice versa) via the explanatory tools of active inference.

## Conclusion

Here, we have argued that active inference, precisely because it is not a theory of consciousness, is uniquely placed among current theoretical approaches to consciousness to do justice to the richness and diversity of phenomena studied under the banner of consciousness science. Crucially, we have argued that active inference is not simply a framework for modelling consciousness. The assumptions implicit in the interpretation of these models — in the context of consciousness science — entail a set of empirical predictions that, if found to be false, would require the revision of the majority of models constructed under active inference. The core of the theory is constituted by an imperative to minimise variational free energy in perception and expected free energy in policy selection, along with the conjecture that the contents of consciousness, including exteroceptive and interoceptive experiences, as well as the state of the brain itself, must — in some way — correspond to the inferred states of the world, at the interface between continuous sensory computations and discrete counterfactual policy selection processes.

The scientific study of consciousness is, ultimately, an empirical science. If active inference is to furnish us with a useful theory of consciousness, then the translation of hard-won theoretical insights into empirical insights must be a priority. At present, there are clear bottlenecks in the translation of abstract models aimed at conceptual clarification into models that provide qualitative explanations of empirical data. And, in turn, translating these more complex theoretical models aimed at qualitative explanation and prediction into minimal models that can be quantitatively compared to data. The majority of explanatory work is still to be done. Nevertheless, we have argued here that the contours of a theory are already beginning to emerge. We are optimistic, therefore, that active inference, as a theory of consciousness, has the resources to eventually provide a minimal set of models that explain and unify the study of conscious contents, self, and state.

## Acknowledgements

The authors thank Mac Shine and members of the Monash M3CS active inference reading group for providing invaluable feedback during the writing of this manuscript. CW is supported by a University of Sydney DVCR strategic postgraduate research scholarship. AWC, JR, and JH are supported by the Three Springs Foundation, and Templeton World Charity Foundation. TP is supported by a NIHR Academic Clinical Fellowship (ref: ACF-2023-13-013). KF is supported by funding for the Wellcome Centre for Human Neuroimaging [203147/Z/16/Z]. AS is supported by the European Research Council (ERC Advanced Investigator Grant 101019254).

### Code and data availability

The novel computational models reported here were implemented with standard model inversion routines (spm_MDP_VB_X.m) available as MATLAB code in the latest version of SPM academic software: http://www.fil.ion.ucl.ac.uk/spm/. Complete code necessary to reproduce the simulations reported in this paper will be made available upon full publication.

### Appendix 1: Levelt's laws under active inference

To show the compatibility of the model of binocular rivalry proposed by Parr, Corcoran et al, (2019) with Levelt's laws, we re-implemented the original model. Briefly, the model had three hidden state factors encoding the identity of the stimulus presented to each eye (with states {left, blank} and {right, blank}) and the locus of attention (with states {left, right}). In other words, one can attend to features consistent with the right stimulus—or to data from the right eye—or to those from the left. The model had two outcome modalities (**A** matrices), one for each eye hidden state factor, which mapped the conditional probability of the observations associated with the stimulus entering each eye to the respective hidden states. The precision of the **A** matrix mapping was conditionally dependent on the attention hidden state factor. When attention was absent the **A** matrices were uniform (e.g. with columns given by $[0.5 \quad 0.5]^T$), and when attention was present the mapping was comparatively precise ($[0.7 \quad 0.3]^T$). This renders the hidden states conditionally independent of unattended observation modalities. The **B** matrices for the hidden state factors associated with each eye were initialised as identity matrices and then passed through a softmax function $s(x_i; \omega) = e^{\boldsymbol{\omega} x_i} / \sum_j^K e^{\boldsymbol{\omega} x_j}$ with precision parameter $\boldsymbol{\omega} = 0.8$ introducing sufficient uncertainty into the agent's transition beliefs to guarantee that in the absence of precise input (i.e. in the absence of attention) uncertainty about hidden states would accumulate over time and the non-attended stimulus would become increasingly epistemically attractive. The **B** matrix for the attention hidden state factor was conditionally dependent on the agent's attention policy {left, right} placing attention transitions under the agent's control. The **C** vector was uniform for both outcome modalities.

To simulate the experimental conditions described by Levelt's propositions we systematically reduced the **A** matrix precision for one (Levelt's second proposition) or both (Levelt's fourth proposition) of the **A** matrix mappings. Levelt's second proposition states that reducing the "strength" (e.g. contrast) of the stimulus entering one eye whilst leaving the strength of the stimulus entering the other eye unaltered should sharply increase of the dominance duration of the percept associated with the unaltered stimulus and lead to a comparatively minor reduction in the dominance duration of the percept associated with the reduced strength stimulus. Consistent with this, reducing the precision of one of the **A** matrices — whilst leaving the other intact — resulted in a sharp increase in the amount of time the stimulus entering the more precise eye was sampled (i.e. attended) and a gradual reduction in the amount of time the stimulus associated with the less precise eye was sampled (**Fig 3b** upper panel). This asymmetry is due to the relative decrease in epistemic value of the less precise stimulus. Levelt's fourth proposition states that increasing the "strength" of the stimulus entering both eyes will result in a reduction in the dominance duration of both stimulus percepts. Consistent with Levelt's fourth proposition, reducing the precision of both **A** matrices increased the amount of time the agent spent sampling the stimulus before switching (**Fig 3b** lower panel). The symmetric reduction in **A** matrix precision drove the agent to spend more time sampling each stimulus to reduce its uncertainty leading to longer dominance durations. Put another way, under reduced **A** matrix precision each stimulus retains its epistemic (i.e. ambiguity reducing) value for longer.

## Appendix 2: Binocular rivalry and the exploration-exploitation dilemma

Based on the trade off between exploration and exploitation (i.e. ambiguity and risk) negotiated by agents minimising expected free energy we sought to generate a novel experimental prediction by extending the simulations of Levelt's propositions described in **appendix 1** to include reward. Specifically, we simulated the conditions described by Levelt's second proposition but in comparison to the first set of simulations which assumed a uniform **C** vector we associated the reduced precision stimulus (e.g. the stimulus with reduced contrast in experimental settings) with reward. At each time step of the simulation the observation associated with the "right" state of the right eye outcome modality was given a value of $\boldsymbol{c}_1 = .25$ leading to an initial bias in the dominance duration for the rewarded stimulus percept when the **A** matrix precision was matched between eyes (i.e. when the epistemic value of each stimulus was matched but risk was lower for the rewarded eye). Importantly, because risk and ambiguity trade off against each other this bias could be compensated for by reducing the precision of the rewarded eye reducing its epistemic (ambiguity reducing) value (**Fig A.2**). We therefore predict that; 1) with matched stimulus strength (contrast) rivalry will be biased to the percept associated with the rewarded stimulus; and 2) that this bias will be compensated for by reducing the epistemic value of the rewarded stimulus. Together, this results in a predicted violation of Levelt's second proposition under reward.

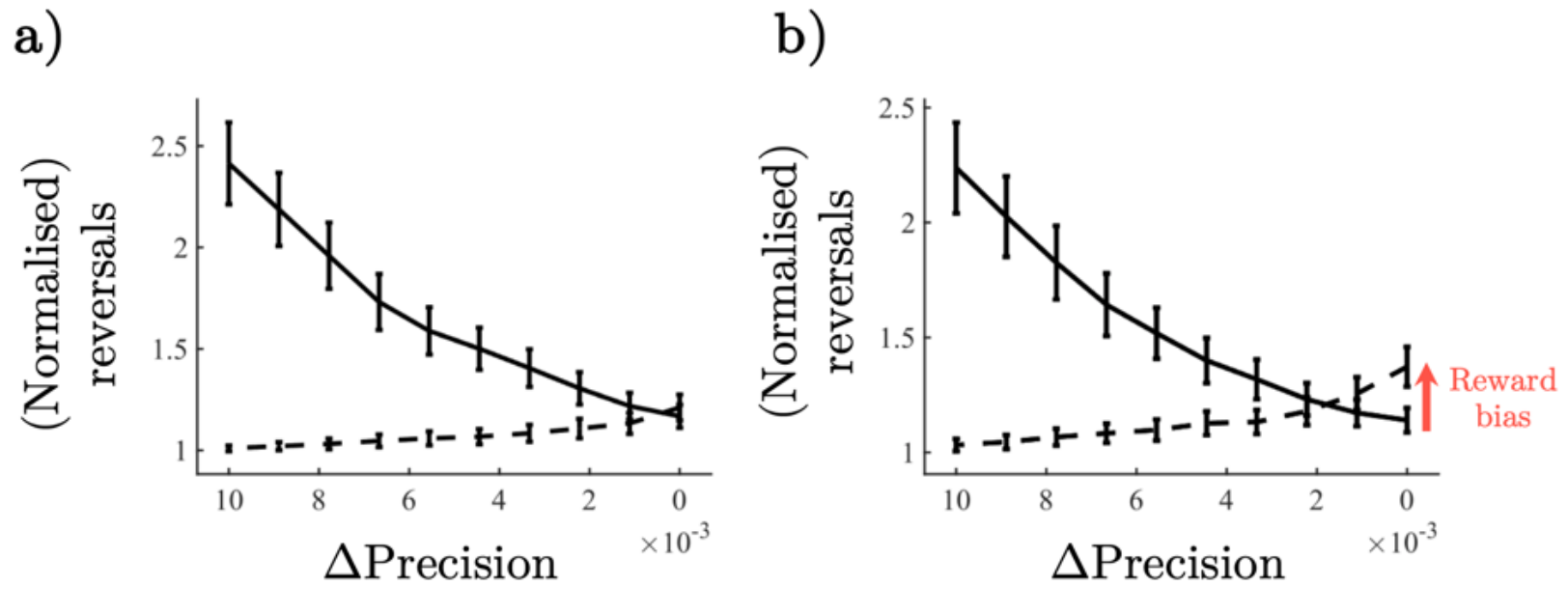

**Figure A2. a)** Simulation of Levelt's second proposition with uniform **C** vector. **b)** Simulation of Levelt's second proposition showing an initial bias to the rewarded stimulus percept (dashed line; $\boldsymbol{c}_1 = .25$) that is compensated for by reducing the precision of the rewarded stimulus.